\documentclass[amssymb,aps,floats,twocolumn,prb,floatfix,showpacs,preprintnumbers]{revtex4-1}
\usepackage[applemac]{inputenc}
\usepackage{graphicx,hyperref}
\DeclareGraphicsExtensions{.eps}

\begin{document}

\title{Ordered moment in the anisotropic and frustrated square
lattice Heisenberg model}

\author{Burkhard Schmidt, Mohammad Siahatgar, and Peter Thalmeier}
\affiliation{Max-Planck-Institut f{\"u}r Chemische Physik fester
Stoffe, 01187 Dresden, Germany}

\begin{abstract}

The two-dimensional frustrated next nearest neighbor Heisenberg model on the
square lattice is a prime example for a spin system where quantum fluctuations
can either destroy or stabilize magnetic order. The phase boundaries and
staggered moment dependence on the frustration ratio $J_2/J_1$ of the exchange
constants are fairly well understood both from approximate analytical and
numerical methods.  In this work we use exact diagonalization for finite
clusters for an extensive investigation of the more general $J_{1a,b}$-$J_2$
model which includes a spatial exchange anisotropy between next-neighbor spins. 
We introduce a systematic way of tiling the square lattice and, for this low
symmetry model, define a controlled procedure for the finite size scaling that
is compatible with the possible magnetic phases.  We obtain ground state
energies, structure factors and ordered moments and compare with the results of
spin wave calculations.  We conclude that $J_{1a,b}$ exchange anisotropy
strongly stabilizes the columnar antiferromagnetic phase for all frustration
parameters, in particular in the region of the spin nematic phase of the
isotropic model.

\end{abstract}


\maketitle

\section{Introduction}
\label{sect:intro}

The two-dimensional (2D) quantum $J_1$-$J_2$ Heisenberg model on a
square lattice has recently found a number of realizations in layered
$V^{4+}$ $(S=1/2)$
compounds~\cite{melzi:00,melzi:01,kaul:04,kaul:05,kini:06}.  By using
thermodynamic~\cite{melzi:00,melzi:01,kaul:04,kaul:05,kini:06} and
high field~\cite{tsirlin:09,tsirlin:09b} investigations as well as
NMR~\cite{melzi:00,melzi:01}, $\mu$SR~\cite{melzi:01}, resonant X-ray
scattering~\cite{bombardi:04} and neutron
diffraction~\cite{bombardi:04,skoulatos:07a,skoulatos:09} it was
possible to locate these compounds in the phase diagram of this model
\cite{misguich:03,shannon:04,shannon:06,schmidt:07b,thalmeier:08}.  It
was concluded that all compounds are lying in the region of the
columnar antiferromagnetic phase corresponding to ordering wave
vector ${\vec Q}=(0,\pi)$ or $(\pi,0)$.

The phase diagram is characterized by a single control parameter, the
frustration angle $\phi=\tan^{-1}(J_2/J_1)$ and the energy scale is fixed by
$J_{\text c}=(J_1^2+J_2^2)^\frac{1}{2}$.  The possible classical phases are a
ferromagnet (FM) with $\pi-\tan^{-1}(1/2)\le\phi\le3\pi/2$, a Néel
antiferromagnet (NAF) with $-\pi/2\le\phi\le\tan^{-1}(1/2)\approx0.15\pi$, and a
columnar antiferromagnet (CAF) with
$\tan^{-1}(1/2)\le\phi\le\pi-\tan^{-1}(1/2)\approx0.85\pi$.  In the transition
regions NAF/CAF and CAF/FM frustration effects destroy the magnetic order in a
small but finite interval.  This can be concluded from exact diagonalization on
small
systems~\cite{dagotto:89,singh:90,schulz:92,richter:93,einarsson:95,schulz:96,zhitomirsky:00,honecker:01,shannon:04,shannon:06,schmidt:07b,thalmeier:08,schmidt:10}
as well as spin wave
calculations~\cite{ceccatto:92,dotsenko:94,shannon:04,schmidt:07b,uhrig:09,schmidt:10},
series expansion~\cite{gelfand:89,oitmaa:96,oitmaa:06}, and large-$N$
expansion~\cite{read:91,sachdev:97}. It has been proposed that the true ground
state in these regions is not a genuine spin liquid but exhibits hidden
(nonmagnetic) order, namely a columnar dimerized phase with an excitation
gap~\cite{singh:99,capriotti:01,capriotti:03,misguich:04} and spin nematic
order~\cite{shannon:06} respectively. The impact of spatial anisotropies on the
columnar dimerized phase around $\phi\approx0.15\pi$ has been studied by exact
diagonalization~\cite{sindzingre:04}, the coupled-cluster
method~\cite{bishop:08}, and density-matrix renormalization group
methods~\cite{moukouri:04}. The model also exhibits anomalies in frustration
dependent magnetic and magnetocaloric
quantities~\cite{schmidt:07b,thalmeier:08}.

The vanadates are in fact not strictly of square lattice type but small
rectangular distortions lead to a small $J_{1a}-J_{1b}$ anisotropy, however it
was shown that it plays only a minor role for these compounds~\cite{tsirlin:09}.
 Furthermore the generalized anisotropic $J_{1a,b}$-$J_2$ Heisenberg model with
large anisotropy has recently been invoked in the interpretation of spin wave
excitations for the Fe-pnictide parent
compounds~\cite{ewings:08,mcqueeney:08,diallo:09,zhao:09}.  It has also been
discussed whether the observation of small ordered moments can be understood
within a frustrated local moment model.  For a discussion of these topics we
refer to Ref.~\onlinecite{schmidt:10} and the numerous references cited therein.
 However the discussion of ordered moment size in Ref.~\onlinecite{schmidt:10}
has so far been exclusively based on approximate analytical spin wave
calculations.

In this paper we want to endeavor a full scale numerical approach based on
exact diagonalization of finite size clusters to clarify the ordered moment size
and its variation with frustration and anisotropy effects in the
$J_{1a,b}$-$J_2$ Heisenberg model by an unbiased method.  In addition to
$J_{\text c}$ and $\phi$ this implies a further parameter $\theta$ (which we
will define later) characterizing the orthorhombic anisotropy.

Our present work has two main objectives.  Firstly it presents a new technical
development how to apply the finite size scaling method systematically to a
Heisenberg model with low spatial symmetry.  In most previous investigations of
the isotropic $J_1$-$J_2$ model either restrictive or not fully systematic
methods have been chosen for the lattice tilings used for the scaling
procedure~\cite{haan:92,schulz:96,betts:99}.  In our case this has to be
generalized because of the lower symmetry of the model and the appearance of
non-degenerate phases with columnar magnetic order.  We will discuss in detail
how all possible tilings can be constructed in a unique way, classify their
symmetry and compatibility with classical ordered phases and introduce a precise
concept of {\em squareness\/} to characterize their usefulness for finite-size
scaling analysis.  In addition we will describe two different methods to
calculate the staggered moment from the correlation functions and discuss their
accuracy.

Secondly, using these new technical developments we will calculate the ground
state energy, structure factor, and staggered moment as functions of frustration
and anisotropy parameters and compare with classical behavior and the results
from spin wave calculations.  In particular we will give a definite answer to
which extent the ordered moment for general frustration and anisotropy
parameters $(\phi,\theta)$ is modified as compared to the isotropic unfrustrated
N\'eel state moment.  We also discuss the influence of the anisotropy on the
stability of the proposed spin nematic phase and show that a tiny deviation from
the isotropic model reestablishes the columnar magnetic order. Further details
on the physcial motivation for the investigations presented in this work are
described extensively in Ref.~\onlinecite{schmidt:10}.

The paper is organized as follows: in Sect.~\ref{sect:model} the model and its
characteristic parameters are defined. In Sect.~\ref{sect:clusters} the
technical implementation of the numerical analysis for the anisotropic model is
described in detail. Then Sect.~\ref{sect:fsscaling} presents the finite-size
scaling procedure, and Sect.~\ref{sect:results} contains a discussion of the
systematic dependence of the thermodynamic limit of the ground state energy
$E_{0}$ and the ordered moment $M_{0}(\vec Q)$ on the model parameters
$(\phi,\theta)$. Finally Sect.~\ref{sect:summary} gives the summary and
conclusions.

\section{The model}
\label{sect:model}

The Hamiltonian for the two-dimensional $S=1/2$ $J_{1a,b}$-$J_2$ model studied
in this work is given by
\begin{equation}
	\mathcal{H} = J_{1a}\sum_{{\langle ij \rangle}_{a}}^N \vec S_i \cdot
	\vec S_j + J_{1b}\sum_{{\langle ij \rangle}_{b} }^N \vec S_i \cdot
	\vec S_j + J_{2}\sum_{{\langle \langle ij \rangle \rangle}}^N \vec
	S_i \cdot \vec S_j.
\label{eqn:ham}
\end{equation}
The first two sums are taken over bonds between nearest-neighbor sites
along the $a$ and $b$ directions of the rectangular lattice,
respectively, and $\langle\langle ij \rangle\rangle$ denotes bonds
connecting the next nearest neighbors along the diagonals of a
rectangular plaquette.  We use a parametrization of the exchange
constants which facilitates the discussion of the whole phase diagram,
and introduce a frustration angle $\phi$ and an anisotropy angle
$\theta$.  With these parameters, we define
\begin{eqnarray}
    J_{1a}&=&\sqrt{2}J_{\text c}\cos\phi\cos\theta,
    \nonumber\\
    J_{1b}&=&\sqrt{2}J_{\text c}\cos\phi\sin\theta,
    \\
    J_{2}&=&J_{\text c}\sin\phi,
    \nonumber\\
    J_{\text c}&=&\sqrt{\frac{1}{2}\left(J_{1a}^{2}+J_{1b}^{2}\right)
    +J_{2}^{2}}.
    \nonumber
    \label{eqn:exndef}
\end{eqnarray}
Again $J_{\text c}$ defines the overall energy scale of the model.

\begin{figure}
    \includegraphics[width=.6\columnwidth]{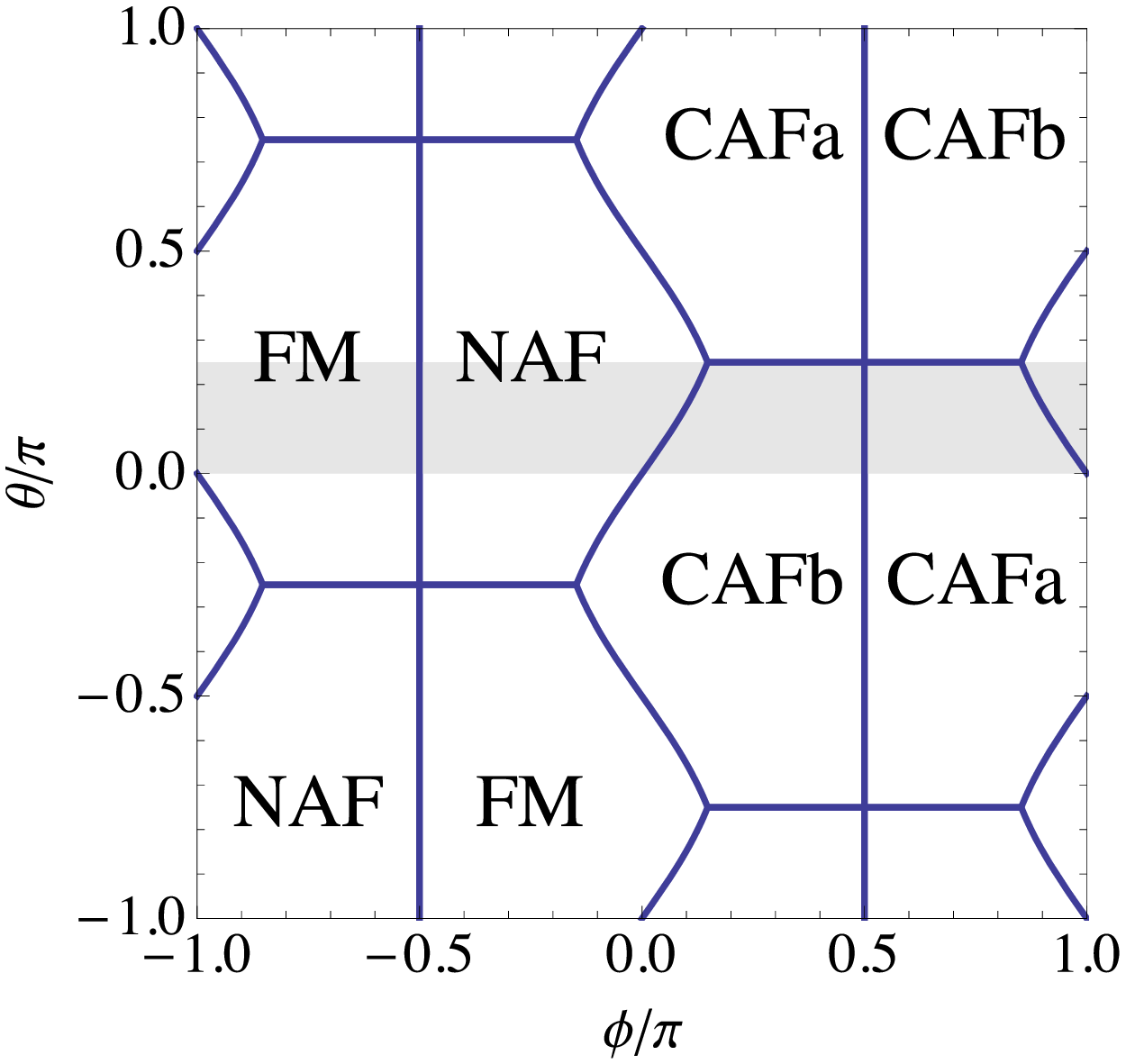}
    \caption{Classical phase diagram of the Hamiltonian defined in
    Eq.~(\ref{eqn:ham}) as a function of the frustration angle $\phi$
    and the anisotropy parameter $\theta$.  For symmetry reasons, it
    is sufficient to restrict the discussion of the phase diagram to
    the parameter range $-\pi\le\phi\le\pi$, $0\le\theta\le\pi/4$
    indicated by the shaded area in the figure.}
    \label{fig:phase}
\end{figure}

The possible classical phases are ferromagnetic (FM, $\vec
Q_{\text{FM}}=(0,0)$), N\'{e}el antiferromagnet (NAF, $\vec Q_{\text{NAF}}
=(\pi,\pi)$) and columnar antiferromagnets (CAFa, $\vec Q_{\text{CAFa}}
=(\pi,0)$, CAFb, $\vec Q_{\text{CAFb}} =(0,\pi)$). The latter are degenerate for
the isotropic ($J_{1a}=J_{1b}$) case with $\theta=\pi/4$ or $\theta=-3\pi/4$.
Fig.~\ref{fig:phase} displays a sketch of the classical phase diagram in the
$\phi$-$\theta$-plane.  The shaded stripe in the plot denotes the parameter
range $-\pi\le\phi\le\pi$, $0\le\theta\le\pi/4$ which can be mapped onto the
whole phase diagram applying discrete symmetry operations under which the
Hamiltonian~(\ref{eqn:ham}) is invariant.

Ref.~\onlinecite{schmidt:10} and references cited therein contain a thorough
discussion of the classical phases and a spin-wave analysis of the ground-state
properties of the Hamiltonian~(\ref{eqn:ham}).  In particular, the inherent
frustration leading to quantum fluctuations and possible moment reduction is
investigated in detail.  It is found that the staggered moment is stabilized in
the columnar phases by introducing a spatial anisotropy, and a strong influence
of frustration on the size of the moment can be excluded.  However, linear
spin-wave analysis a priori is a semiclassical theory, and naturally the
question arises to what extent these results can be confirmed by strictly
quantum-mechanical methods. An unbiased but technically more involved
approach is to determine the ground-state properties of the
Hamiltonian~(\ref{eqn:ham}) by exact diagonalization on small finite tiles with
periodic boundary conditions and to extrapolate the results to the thermodynamic
limit using a scaling analysis.

\section{Technical prerequisites for a finite-size scaling analysis}
\label{sect:clusters}

The most difficult aspect of finite size cluster calculations is the
implementation of an efficient finite size scaling procedure to obtain
reliable values of physical quantities in the thermodynamic limit.  In
this section we present in detail the necessary ingredients.  We
describe how we generate lattice tilings used for exact
diagonalization, classify them according to a newly introduced concept
of their compactness or squareness (which will be defined later),
their point-group symmetry, and their compatibility with the classical
phases of our model.  These properties allow us to discriminate
systematically the various clusters according to their usefulness for
the finite size scaling analysis which is of central importance to
obtain reliable results.  Furthermore we will discuss the derivation
of the set of wave vectors associated with a given tile and describe
two different finite-size scaling procedures to calculate the ordered
moment.

\subsection{Tiling the square lattice}

For a finite-size scaling analysis, it is common practice to select
tiles according to certain geometrical or topological properties.  We
first briefly describe three different schemes applied in the past to
$S=1/2$ spin models before turning to the more general selection 
scheme used in this paper.

Haan et al.~\cite{haan:92} discuss the $S=1/2$ nearest-neighbor
Heisenberg antiferromagnet with helical boundary conditions and define
an asymmetry parameter $A=|\ell_{1}-\ell_{2}|/(\ell_{1}+\ell_{2})$,
where $\ell_{i}$ are the lengths of the edge vectors of the tiles
under consideration.  A square-shaped tile (considered as ``good'')
has $A=0$, but this is true for general diamond-shaped tiles, too.
Those tiles having ``small $A$'' are selected for scaling.

Restricting to strictly square-shaped tiles having at least ${\cal
C}_{4}$ point-group symmetry is the recipe used by Schulz et
al.~\cite{schulz:96} for discussing ground-state energy and ordered
moment of the $S=1/2$ antiferromagnetic $J_{1}-J_{2}$ model.  However,
with this criterion only very few (two to four) tiles are eventually
used for a linear least-squares fit.

Another approach can be found in Ref.~\onlinecite{betts:99},
discussing the $S=1/2$ nearest-neighbor $XY$ and Heisenberg
antiferromagnets with periodic boundary conditions.  The authors
introduce a parameter called the {\em topological imperfection\/}
of a tile, where a topologically perfect tile is defined as follows: A
given lattice point on a tile contains $n_{1}$ nearest neighbors,
$n_{2}$ next-nearest neighbors, and so on, up to the
$i_{\text{max}}$th-nearest neighbors where the sum over the $n_{i}$
reaches the tile area $N$.  (Distance is measured as the minimal
number of hops needed to get from one point to another.)  If for all
$i<i_{\text{max}}$ we have $n_{i}=4i$, which is the number of
$i$th-nearest neighbors on the infinite lattice, a tile is considered
as topologically perfect.  This concept is then extended to the
notion of topologically perfect bipartite Néel lattices, i.\,e., the
same conditions as described above are applied individually to each of
the two sublattices for antiferromagnetic Néel order.  However, tiles
are eventually chosen by hand in order to achieve a smooth finite-size
scaling behavior of the ground-state energy per site and the square of
the magnetization or staggered moment, respectively.

The examples given above are in no way exhaustive, but illustrate one
problem common to any finite-size scaling analysis, which becomes
particularly apparent when applied to the full phase diagram of the
$J_{1a,b}$-$J_{2}$ model.  Firstly, only very few tilings might
survive the final selection, making a linear two-parameter $\chi^{2}$
fit to the ground-state energy or squared ordered moment of the
$J_{1a,b}$-$J_{2}$ model questionable, not to speak about higher-order
correction terms included in the fitting 
procedure.~\cite{hasenfratz:93}

Secondly, the selection contains some arbitrariness which in our case
would lead to selecting different tiles for scaling for different sets
of exchange parameters, even within the same classical phase.

Thirdly, given the edge vectors $\vec a_{i}$ of a particular
parallelogram, this is not a unique way to tile the square lattice.
For example, upon replacing $\vec a_{1}$ by, say $2\vec a_{2}-\vec
a_{1}$, we get a new tile with identical area which leads to an
identical structure of the resulting torus when introducing periodic
boundary conditions.  Even worse: In general, there are many possible
generator matrices $M=(\vec a_{1},\vec a_{2})$ of the same lattice
tiling $\Lambda_{M}$.

However a more systematic approach to select tilings is possible.  We
will describe this scheme here and employ it for the
$J_{1a,b}$-$J_{2}$ model.

A basic requirement is a unique description of the lattice
tilings.~\cite{schrijver:86,domich:87,lyness:91,stewart:97} To achieve this, we
need some definitions.  First, we introduce {\em unimodular\/} matrices $U$,
which are defined as integer matrices of dimension $s$ with determinant $\pm1$. 
Let us further denote with $U^{(i,j)}$ a unit matrix modified by having a single
additional nonzero unit element at position $(i,j)$, and introduce $S^{(i)}$ as
a unit matrix modified by having the $(i,i)$th element replaced by $-1$.  The
unimodular matrices of dimension $s$ form a (non-abelian) group $U_{\text s}$
under matrix multiplication, and the generators of this group can be the
elements $U^{(i,j)}$ and $S^{(i)}$.

We need another definition: A $s\times s$ integer matrix
$H=\left(h_{ij}\right)$ is in {\em Hermite Normal Form\/} (HNF) if and
only if
\begin{eqnarray}
    h_{ii}\ge1,&\quad&i=1\ldots s,
    \nonumber\\
    h_{ij}=0,&\quad&1\le j<i\le s,
    \label{eqn:hnfdef}
    \\
    h_{ij}\in\left[0,h_{ii}\right),&\quad&1\le i<j\le s.
    \nonumber
\end{eqnarray}
It can be shown~\cite{schrijver:86} that (a) an arbitrary nonsingular
integer matrix $M$ can be uniquely represented by a unimodular
matrix $U$ and an HNF matrix $H$ such that
\begin{equation}
    U\cdot M=H,
    \label{eqn:umh}
\end{equation}
and (b) that for two HNF matrices $H$ and $H'$ generating the lattice
tilings $\Lambda_{H}$ and $\Lambda_{H'}$, we have
\begin{equation}
    \Lambda_{H}\equiv\Lambda_{H'}\leftrightarrow
    H=H'.
\end{equation}
In this way, we can uniquely represent an arbitrary lattice tiling of 
any given primitive Bravais lattice.

The two-dimensional HNF matrices $H$ have the form
\begin{equation}
    H=\left(
    \begin{array}{cc}
        h_{11}&h_{12}\\
        0&h_{22}
    \end{array}
    \right)
    \label{eqn:hnf2}
\end{equation}
representing tiles with the special edge vectors $\vec
h_{1}=\left(h_{11},h_{12}\right)$ and $\vec
h_{2}=\left(0,h_{22}\right)$ and area or number of sites
$N=h_{11}h_{22}$.  For numerical purposes, it would already be
sufficient to
implement an algorithm using only Eqs.~(\ref{eqn:hnfdef})
and~(\ref{eqn:hnf2}) for building the list of possible tilings. 

\subsection{Compact tiles}
\label{sect:squareness}

In the spirit of Ref.~\onlinecite{schulz:96}, where only square-shaped
tiles have been used, we want to generalize this concept and utilize
the notion of ``squareness'' or ``compactness'' of a tile for
selecting the proper tiles for finite-size scaling.  For this purpose,
we introduce a parameter
\begin{eqnarray}
    \rho(M)&=&
    \frac{\left|\mathop{\rm 
    Det}M\right|}{\left|\left|M\right|\right|},
    \nonumber\\
    \left|\left|M\right|\right|&=&
    \left(\frac{1}{s}
    \sum_{i=1}^{s}\sqrt{\mathop{\rm Det}\left(M_{i}^{T}M_{i}\right)}
    \right)^{s/(s-1)}
    \label{eqn:squareness}
\end{eqnarray}
for a nonsingular integer $s\times s$ matrix $M$, where $M_{i}$ is the
(non-square) matrix formed by dropping the $i$th row of $M$.
$\rho(M)$ measures the ``compactness'' (``squareness'' in two spatial
dimensions) of the $s$-dimensional parallelotope spanned by the row
vectors of $M$: We have $0<\rho(M)\le1$, and $\rho(M)=1$ if and only
if $M$ describes an $s$-dimensional cube, which we regard as the most
compact lattice tiling in dimension $s$.

However, calculating $\rho(H)$ for the HNF representation of a lattice
tiling is not very useful: According to Eq.~(\ref{eqn:umh}), a single
HNF matrix $H$ represents a whole class ${\cal C}_{H}$ of tiles with
matrix representation $M$ which all lead to an identical lattice
tiling $\Lambda_{H}=\Lambda_{M}$.  But in general, two matrices $M\ne
M'$ with $\Lambda_{M}=\Lambda_{M'}$ have $\rho(M)\ne\rho(M')$.  From
each ${\cal C}_{H}$, we therefore have to choose a tile which has the
maximum compactness of all tiles of its class,
\begin{equation}
    M_{\text c}=M:\rho(M)=\max_{{\cal T}\in{\cal 
    C}_{H}}\left(\rho(M_{{\cal T}})\right)
\end{equation}
and assign $\rho(M_{\text c})$ to the HNF representant $H=UM_{\text 
c}$ of its class ${\cal C}_{H}$.

With this scheme, we find 816 different classes of tiles with size $N$ between
$8$ and $32$.  It is impossible to list all of them in this article.  Instead,
to illustrate the principle we display in Fig.~\ref{fig:hnf8} of
Appendix~\ref{app:tiling} a list of all possible distinct ways to tile the
square lattice using tiles of just the smallest useful size $N=8$.

To label the tiles in a unique way, we use the scheme
$N$:$h_{11}$-$h_{12}$, where $N$ is the number of sites or tile area
and $h_{ij}$ are the coordinates of the first edge vector of the HNF
representation of a tile.  The eight-site square for example has the
label $8$:$2$-$2$ (Fig.~\ref{fig:hnf8}, third from below).

\subsection{Construction of the Brillouin zone for a finite lattice 
tiling}

To reduce the size of the matrices representing the Hamiltonian, it is useful to
work with a basis reflecting the periodicity of the lattice, i.\,e., we
construct Bloch states from the states of the $(N,S_{z})$ basis by applying
Bloch's theorem. Algorithmically, the states of the original spinor product
basis are collected into classes of wave functions which related to each other
by translations with translation vector $\vec r$ and associated phase factor
${\rm e}^{{\rm i}\vec k\vec r}$.

For the finite tilings we are using, the possible wave vectors $\vec k$ can
assume only certain values. In this section, we discuss how to determine these
wave vectors which are contained in the (first) Brillouin zone for a given
lattice tiling.  We use the fact that a translation by a reciprocal lattice
vector will not change the phase of a wave function, as required by Bloch's
theorem.  We construct the first Brillouin zone such that the origin is located
in one of its corners.  This is different from the commonly used Wigner-Seitz
construction for infinite lattices, where the origin is in the center of the
Brillouin zone.  We cannot apply the Wigner-Seitz construction directly to an
arbitrary finite lattice, because we require the wave vector $\vec k=0$ to be
part of the set of reciprocal lattice points generated, and the geometrical
center of a given reciprocal lattice tile constructed as described here does not
necessarily have a wave vector associated with it.

For the edge vectors $\vec a_{1}$ and $\vec a_{2}$ of a tile and the
corresponding basis vectors $\vec b_{1}$ and $\vec b_{2}$ of the reciprocal
lattice, we get from the orthogonality condition
\begin{equation}
    \vec b_{1}=\frac{2\pi}{N}\left(a_{22}\atop{-a_{21}}\right),
    \quad
    \vec b_{2}=\frac{2\pi}{N}\left({-a_{12}}\atop a_{11}\right)
    \label{eqn:b1andb2}
\end{equation}
with integer coefficients $a_{ij}$, where
\begin{equation}
    N=\left|\mathop{\rm Det}\left(\vec a_{1},\vec a_{2}\right)\right|
\end{equation}
is the number of sites or area of the tile under consideration. The reciprocal
lattice vectors $\vec G_{1}=2\pi\vec e_{x}$ and $\vec
G_{2}=2\pi\vec e_{y}$, where $\vec e_{x,y}$ are the Cartesian unit
vectors, are given by
\begin{equation}
    \vec G_{i}=a_{1i}\vec b_{1}+a_{2i}\vec b_{2},
    \quad i=1,2
    \label{eqn:gbi}
\end{equation}
in terms of the reciprocal basis vectors.  Within this coordinate
system, the reciprocal lattice vectors $\vec G_{1}$ and $\vec G_{2}$
span the parallelogram making up the first Brillouin zone, which
contains exactly $N$ wave vectors $\vec k=k_{1}\vec b_{1}+k_{2}\vec
b_{2}$ with integer coefficients $k_{i}$.

\begin{figure}
    \centering
    \null\hfill
    \includegraphics[width=.49\columnwidth]{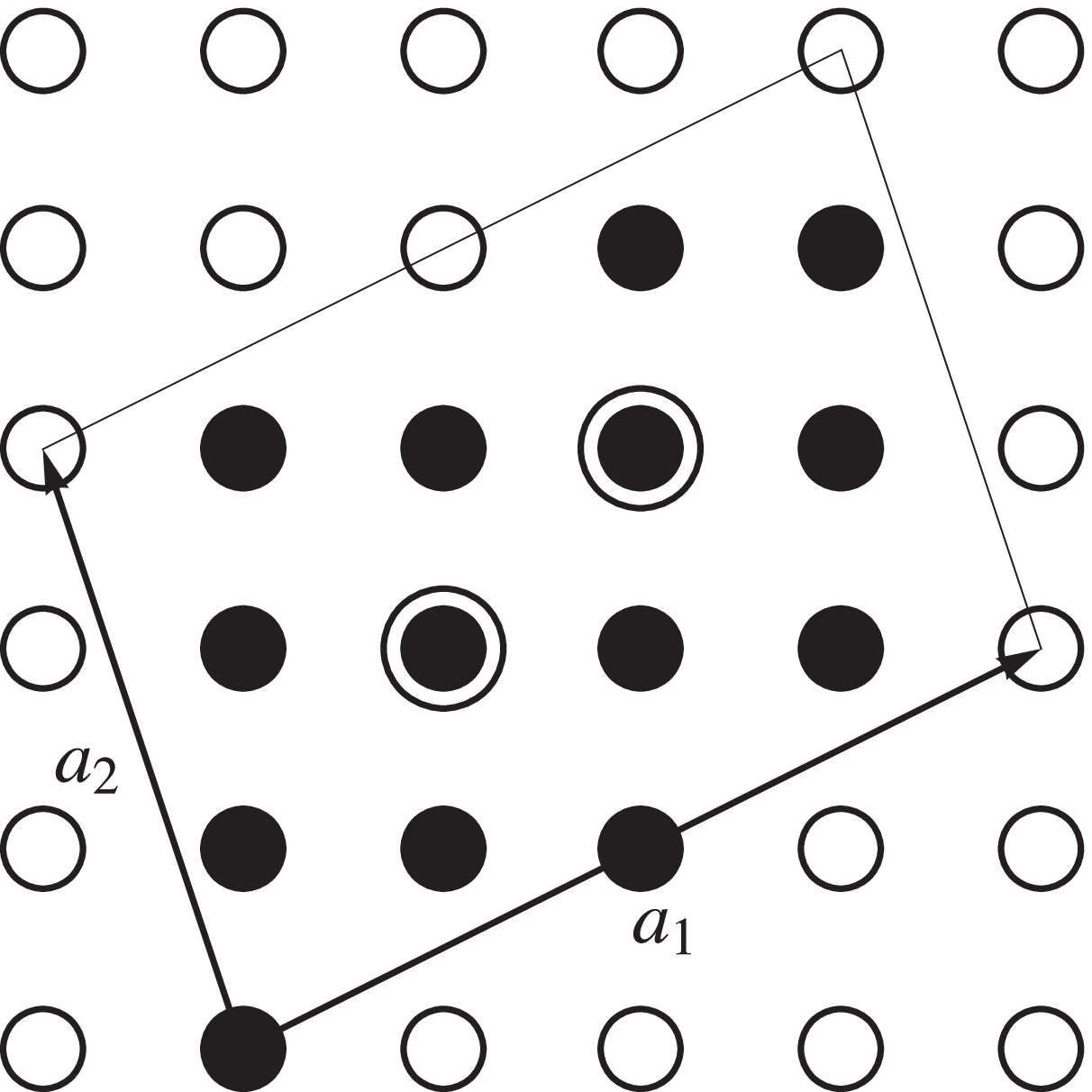}
    \hfill
    \includegraphics[width=.49\columnwidth]{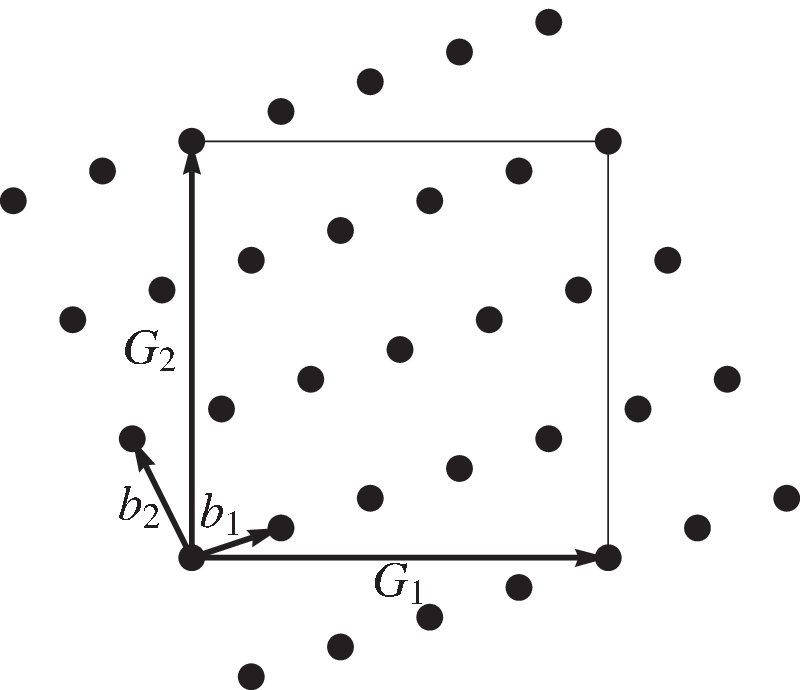}
    \hfill\null \caption{Tile 14:1-11 in direct space (left), and the
    corresponding reciprocal lattice (right) according to
    Eqs.~(\ref{eqn:b1andb2}) and~(\ref{eqn:gbi}).  The two circles on
    the left-hand side mark the two equivalent points of the tile with
    maximum distance to the origin.  Also shown is the first Brillouin
    zone of the lattice with edge vectors (reciprocal lattice vectors)
    $\vec G_{1}$ and $\vec G_{2}$.}
    \label{fig:coords14}
\end{figure}

These wave vectors can be found with the following criterion: The projections of
$\vec k$ onto the reciprocal lattice vectors $\vec G_{i}$ must be positive
semidefinite (we put the origin $(k_{1},k_{2})=(0,0)$ into the lower left corner
of the Brillouin zone) and less than the length of the latter, i.\,e., $0\le\vec
k\cdot\vec G_{i}<\vec G_{i}\cdot\vec G_{i}=4\pi^{2}$.  With the expressions
above for the basis vectors $\vec b_{i}$ and the reciprocal lattice vectors
$\vec G_{i}$, we find as our defining relations for possible wave vectors $\vec
k$
\begin{equation}
    0\le\mathop{\rm Det}(\vec k,\vec G_{2})<N
    \ \wedge\ 
    0\le-\mathop{\rm Det}(\vec k,\vec G_{1})<N,
    \label{eqn:cond}
\end{equation}
writing the column vectors $\vec G_{i}$ and $\vec k$ in the coordinate
system spanned by the basis vectors $\vec b_{i}$.
Conditions~(\ref{eqn:cond}) specify exactly $N$ wave vectors $\vec
k=k_{1}\vec b_{1}+k_{2}\vec b_{2}$. Fig.~\ref{fig:coords14} 
illustrates this procedure for tile number 14:1-11.

\subsection{Space group symmetry}
\label{sect:spacegroups}

\begin{figure}
    \centering
    \includegraphics[width=.6\columnwidth]{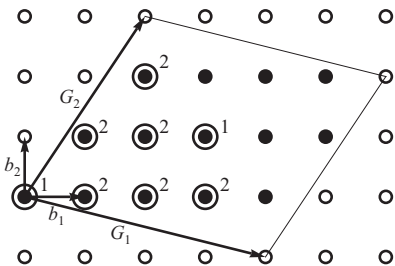}
    \caption{Reciprocal lattice tile 14:1-11.  The graphics shows the
    same tile as in the right-hand side of Fig.~\ref{fig:coords14}
    corresponding to the first Brillouin zone, but here in the
    coordinate system spanned by the reciprocal basis vectors $\vec
    b_{1}$ and $\vec b_{2}$, indicated by the short arrows.  The
    dotted circles denote the irreducible wedge of the Brillouin zone (tile 14:1-11 has ${\cal
    C}_{2}$ symmetry only), numbers give the size of the star of the
    corresponding wave vector.}
    \label{fig:rlocal}
\end{figure}

To avoid unnecessary computations, we calculate Lanczos eigensystems only at
those wave vectors $\vec k$ which are not related by a space group operation. 
Therefore we have to determine the irreducible wedge of the Brillouin zone. 
Figure~\ref{fig:rlocal} illustrates this for tile 14:1-11, which has ${\cal
C}_2$ point group symmetry: The number of wave vectors is reduced from a total
of 14 to eight nonequivalent ones.  The small numbers in the figure denotes the
size of the star of the corresponding wave vector.

Since we are working with mappings of the full tile, it is useful to
work in a coordinate system of the reciprocal lattice vectors $\vec
G_{1}$ and $\vec G_{2}$.  For the projection of an arbitrary wave
vector $\vec k=\hat k_{1}\vec G_{1}+\hat k_{2}\vec G_{2}$ (not
necessarily located in the first Brillouin zone) onto the reciprocal
lattice vectors, we have
\begin{eqnarray}
    \hat k_{1}&=&
    \frac{\vec k\cdot\vec G_{1}}{|\vec G_{1}|^{2}}
    =
    \frac{1}{N}\mathop{\rm Det}\left(\vec k,\vec G_{2}\right),
    \\
    \hat k_{2}&=&
    \frac{\vec k\cdot\vec G_{2}}{|\vec G_{2}|^{2}}
    =
    -\frac{1}{N}\mathop{\rm Det}\left(\vec k,\vec G_{1}\right)
\end{eqnarray}
(compare Eq.~(\ref{eqn:cond})).  Note the minus sign in the last
equation and the interchange of $\vec G_{1}$ and $\vec G_{2}$: The
projection
of $\vec k$ onto $\vec G_{1}$ is proportional to the area of the
parallelogram spanned by $\vec k$ and $\vec G_{2}$ and vice versa.

We can write
\begin{equation}
    \hat k_{i}=\frac{n_{i}}{N}+m_{i},\quad 0\le n_{i}<N,
    \quad m_{i}\in\mathbb{Z}
\end{equation}
for the coefficients of a wave vector $\vec k=M\cdot\vec q$ resulting
from the application of a space group operation $M$ onto a wave vector
$\vec q$ in the first Brillouin zone.  Mapping $\vec k$ back into the
first Brillouin zone amounts to setting $m_{i}=0$ in the equation
above.  This gives us a convenient and numerically well defined
procedure for mapping the Brillouin zone onto its irreducible part.

\subsection{Ordered ground states}
\label{sect:ogs}

The classical $J_{1a,b}$-$J_{2}$ model on the square lattice has four
ground states with ordering wave vectors $\vec Q=(0,0)$, $(\pi,\pi)$,
and $\vec Q=(\pi,0)$ or $(0,\pi)$.  Although in the quantum case the
corresponding wave
functions, except for the ferromagnet, are not eigenstates of the
Hamiltonian, it is important that the tilings of the infinite
lattice are chosen such that these states corresponding to the
classically ordered phases are not suppressed when applying periodic
boundary conditions.

We can determine those tiles compatible with a classical ground state 
by applying a test for the existence of the corresponding classical 
ordering vector $\vec Q$ in the list of wave vectors for a given tile,
\begin{equation}
    n_{1}\vec b_{1}+n_{2}\vec b_{2}=\vec Q,\quad n_{i}\in\mathbb{Z}.
\end{equation}
From Eqs.~(\ref{eqn:b1andb2}), we get
\begin{equation}
    \frac{2\pi}{N}\left[
    n_{1}\left(a_{22}\atop{-a_{21}}\right)+
    n_{2}\left({a_{12}}\atop a_{11}\right)
    \right]=
    \left(Q_{1}\atop Q_{2}\right),
    \label{eqn:qtest}
\end{equation}
which has to be fulfilled for integer coefficients $n_{i}$.

\begin{itemize}
    \item Ferromagnet: $\vec Q_{\text{FM}}=(0,0)$. Since we have chosen the
    phase of our wave functions such that $\vec k=0$ is always a valid wave 
    vector, all tiles are compatible with the ferromagnet.
    
    \item CAFa phase: $\vec Q_{\text{CAFa}}=(\pi,0)$, and we have
    \begin{equation}
        n_{1}=\frac{a_{11}}{2},\quad
        n_{2}=\frac{a_{21}}{2}.
    \end{equation}
    Stated physically, the Mannheim distance (or Manhattan distance) $d_{\text
    M}(\vec p_{1},\vec p_{2})=\sum_{i}\left|p_{1i}-p_{2i}\right|$, counting the minimal
    number of hops on the lattice needed to get from one point to
    another, of any two points $\vec p_{i}$ and $\vec p_{j}$ belonging
    to the same sublattice projected onto the $x$ direction of the
    lattice must be even,
    \begin{equation}
	d_{x}(\vec p_{i},\vec p_{j})=|p_{jx}-p_{ix}|
	\equiv 2n,\  n\in\mathbb{N}_{0}.
    \end{equation}
    \null
    
    \item CAFb phase: $\vec Q_{\text{CAFb}}=(0,\pi)$. We get from 
    condition~(\ref{eqn:qtest})
        \begin{equation}
        n_{1}=\frac{a_{12}}{2},\quad
        n_{2}=\frac{a_{22}}{2},
    \end{equation}
    so the components $a_{12}$ and $a_{22}$ parallel to $\vec e_{y}$
    must be even numbers.  For completeness, the appropriate 
    condition is given by
    \begin{equation}
	d_{y}(\vec p_{i},\vec p_{j})
	\equiv 2n,\  n\in\mathbb{N}_{0}
    \end{equation}
    for any two lattice points $\vec p_{i}$ and $\vec p_{j}$ with
    $S_{i}^{z}=S_{j}^{z}$.
        
    \item N\'{e}el phase: $\vec Q_{\text{NAF}}=(\pi,\pi)$. Eq.~(\ref{eqn:qtest})
can 
    be solved with
    \begin{equation}
        n_{1}=\frac{a_{11}+a_{22}}{2},\quad
        n_{2}=\frac{a_{21}+a_{22}}{2}.
    \end{equation}
    This is equivalent to
    \begin{equation}
	d_{\text M}(\vec p_{i},\vec p_{j})
	\equiv 2n,\  n\in\mathbb{N}_{0}
    \end{equation}
    for any two lattice points $\vec p_{i}$ and $\vec p_{j}$ on the 
    \emph{same} sublattice.
    
    \item All four phases: All components of the edge vectors $\vec
    a_{1}$ and $\vec a_{2}$ must be even individually in order to be
    compatible with the full set of classical phases.
\end{itemize}
In this way, we can classify any given lattice tiling with respect to 
the classical phases it belongs to.
    
\subsection{Selection of tiles}

\begin{table*}
    \centering
    \[
\begin{array}{cccccccccc}
	    \text{Tile} & \text{NAF} & \text{CAFa} & \text{CAFb} & 
            \Box &
            {\cal C}_{2} & {\cal C}_{2V\,\text{rect}} &
	    {\cal C}_{2V\,\text{dia}} &
	    {\cal C}_{4} & {\cal C}_{4V} \\
	    \hline
 \text{\bf\underline{8:2-2}} & \bullet  & \bullet  & \bullet  &
\text{ 1.000} & \bullet  & \bullet  & \bullet  & \bullet  & \bullet
\\
 \hline
 \text{10:1-3} & \bullet  & - & - & \text{ 1.000} & \bullet  & - & -
& \bullet  & - \\
 \text{10:1-4} & - & - & \bullet  & \text{ 0.966} & \bullet  & - & -
& - & - \\
 \text{10:2-2} & - & \bullet  & - & \text{ 0.966} & \bullet  & - & -
& - & - \\
 \hline
 \text{12:3-0} & - & - & \bullet  & \text{ 0.980} & \bullet  &
\bullet  & - & - & - \\
 \text{12:4-0} & - & \bullet  & - & \text{ 0.980} & \bullet  &
\bullet  & - & - & - \\
 \text{12:1-5} & \bullet  & - & - & \text{ 0.960} & \bullet  & - &
\bullet  & - & - \\
 \text{\bf\underline{12:2-2}} & \bullet  & \bullet  & \bullet  &
\text{ 0.901} & \bullet  & - & - & - & - \\
 \hline
 \text{14:1-3} & \bullet  & - & - & \text{ 0.961} & \bullet  & - & -
& - & - \\
 \text{14:1-4} & - & - & \bullet  & \text{ 0.938} & \bullet  & - & -
& - & - \\
 \text{14:2-3} & - & \bullet  & - & \text{ 0.938} & \bullet  & - & -
& - & - \\
 \hline
 \text{\bf\underline{16:4-0}} & \bullet  & \bullet  & \bullet  &
\text{ 1.000} & \bullet  & \bullet  & \bullet  & \bullet  & \bullet
\\
 \hline
 \text{18:3-3} & \bullet  & - & - & \text{ 1.000} & \bullet  &
\bullet  & \bullet  & \bullet  & \bullet  \\
 \text{18:1-4} & - & - & \bullet  & \text{ 0.975} & \bullet  & - & -
& - & - \\
 \text{18:2-4} & - & \bullet  & - & \text{ 0.975} & \bullet  & - & -
& - & - \\
 \hline
 \text{\bf\underline{20:2-4}} & \bullet  & \bullet  & \bullet  &
\text{ 1.000} & \bullet  & - & - & \bullet  & - \\
 \hline
 \text{22:1-6} & - & - & \bullet  & \text{ 0.981} & \bullet  & - & -
& - & - \\
 \text{22:2-4} & - & \bullet  & - & \text{ 0.981} & \bullet  & - & -
& - & - \\
 \text{22:1-5} & \bullet  & - & - & \text{ 0.961} & \bullet  & - & -
& - & - \\
 \hline
 \text{24:1-10} & - & - & \bullet  & \text{ 0.988} & \bullet  & - & -
& - & - \\
 \text{24:2-5} & - & \bullet  & - & \text{ 0.988} & \bullet  & - & -
& - & - \\
 \text{24:1-7} & \bullet  & - & - & \text{ 0.980} & \bullet  & - &
\bullet  & - & - \\
 \text{\bf\underline{24:4-0}} & \bullet  & \bullet  & \bullet  &
\text{ 0.960} & \bullet  & \bullet  & - & - & - \\
 \hline
 \text{26:1-5} & \bullet  & - & - & \text{ 1.000} & \bullet  & - & -
& \bullet  & - \\
 \text{26:1-10} & - & - & \bullet  & \text{ 0.964} & \bullet  & - & -
& - & - \\
 \text{26:2-5} & - & \bullet  & - & \text{ 0.964} & \bullet  & - & -
& - & - \\
 \hline
 \text{28:1-8} & - & - & \bullet  & \text{ 0.986} & \bullet  & - & -
& - & - \\
 \text{28:4-3} & - & \bullet  & - & \text{ 0.986} & \bullet  & - & -
& - & - \\
 \text{\bf\underline{28:2-4}} & \bullet  & \bullet  & \bullet  &
\text{ 0.961} & \bullet  & - & - & - & - \\
 \hline
 \text{30:5-0} & - & - & \bullet  & \text{ 0.992} & \bullet  &
\bullet  & - & - & - \\
 \text{30:6-0} & - & \bullet  & - & \text{ 0.992} & \bullet  &
\bullet  & - & - & - \\
 \text{30:1-5} & \bullet  & - & - & \text{ 0.974} & \bullet  & - & -
& - & - \\
 \hline
 \text{\bf\underline{32:4-4}} & \bullet  & \bullet  & \bullet  &
\text{ 1.000} & \bullet  & \bullet  & \bullet  & \bullet  & \bullet 
\end{array}
    \]
    \caption{Label, classical phase compatibility, squareness, and
    point groups for selected lattice tilings between 8 and 32 sites.
    For each even area $N$ and for each classical phase, the list
    contains the compatible tile with maximum squareness as defined in
    Eq.~(\ref{eqn:squareness}).  For $N=12$ and~$24$, the tiles
    compatible with all classical phases are included, too, although
    they have a comparatively small squareness.  The tile labels have
    the form $N$:$h_{11}$-$h_{12}$, where $h_{1j}$ are the components
    of the first edge vector of a tile in HNF representation.  Those
    tiles compatible with all four classical phases, required for the
    discussion of the spatially isotropic model with columnar order,
    are underlined and typeset in bold.}
    \label{tbl:pg}
\end{table*}

In order to discuss spatial anisotropies, we do not regard lattice tilings as
equivalent which are related by a point-group operation on the square lattice.
An example would be the four different tiles 8:1-2, 8:1-6, 8:2-1, and 8:2-3, see
Fig.~\ref{fig:hnf8}. In this way, we get $816$ different tilings of the square
lattice with tiles between $8$ and $32$ sites. Out of these, we select, for each
even tile area and for each classical phase, the tile having the maximum
squareness.  In addition, for $N=4\ell$, $\ell\in\mathbb{N}$, we include the
tile with maximum squareness containing both CAFa and CAFb ordering vectors,
which is equivalent to containing all four classical ordering vectors.  These
special tiles are of particular importance for the spatially isotropic model,
due to the degeneracy of the two columnar phases in this case.  The resulting
list is displayed in Table~\ref{tbl:pg}. Each line contains the tile label, its
compatibility with classical phases (Sect.~\ref{sect:ogs}), its squareness
(Sect.~\ref{sect:squareness}), and its point group symmetry
(Sect.~\ref{sect:spacegroups}).

For the ${\cal C}_{2V}$ group, two sets of mirror ``planes'' exists:
The point group ${\cal C}_{4V}$ contains two isomorphic subgroups
${\cal C}_{2V\,\text{rect}}$ and ${\cal C}_{2V\,\text{dia}}$,
corresponding to a tile with either a rectangular shape (mirror planes
parallel to the edges) or a diamond-like shape (mirror planes along
the
diagonals).  This distinction is important when discussing
orthorhombic or trigonal symmetry breaking, since ${\cal C}_{4V}$ is
reduced to the respective ${\cal C}_{2V}$ symmetry in this case.  For
simplicity, we also label those tiles having ${\cal C}_{2V}$ but not
${\cal C}_{4V}$ symmetry accordingly in Table~\ref{tbl:pg}.

\subsection{Static structure factor and ordered moment}
\label{sect:structurefactor}

In a finite system at zero magnetic field the ground state has $S_z=0$
and therefore does not exhibit the spontaneous symmetry breaking of
the infinite lattice.  The ordered moment $M(\vec Q)$ rather has to be
obtained indirectly from the properly normalized static structure
factor according to

\begin{eqnarray}
    S_{N}(\vec Q)&=&\frac{1}{\cal N}\sum_{i,j=1}^{N}
    \left\langle\vec S_{i}\vec S_{j}\right\rangle
    {\rm e}^{{\rm i}\vec Q\left(\vec R_{i}-\vec R_{j}\right)}
    \label{eqn:sf}
    \\
    &=&
    \frac{N}{\cal N}
    \left[
    \left\langle\vec S_{1}\vec S_{1}\right\rangle+
    \sum_{j=2}^{N}
    \left\langle\vec S_{1}\vec S_{j}\right\rangle
    {\rm e}^{{\rm i}\vec Q\left(\vec R_{1}-\vec R_{j}\right)}
    \right],
    \nonumber
\end{eqnarray}
where the angular brackets denote the ground-state expectation value
in the $S_{z}=0$, $\vec k=0$ subspace of the Hilbert space.  In the
thermodynamic limit, if $\vec Q$ is the ordering vector of the
corresponding classical phase, we can then identify
\begin{equation}
    M^{2}(\vec Q)=\lim_{N\to\infty}M^{2}_{N}(\vec Q)
    =\zeta(\vec Q)\lim_{N\to\infty}S_{N}(\vec Q)
    \label{eq:Mfirst}
\end{equation}
with the appropriate normalization $\cal N$ of $S_{N}(\vec Q)$. Here 
we have introduced a factor
\begin{equation}
    \zeta(\vec Q)=\left\{
    \begin{array}{cl}
        1,&\vec Q=0\ \mbox{or}\ (\pi,\pi)
        \\
        2,&\vec Q=(\pi,0)\ \mbox{or}\ (0,\pi)
    \end{array}
    \right.
\end{equation}
to account for the additional lattice rotation symmetry breaking in
the CAFa and CAFb phases.~\cite{schulz:96}

For a perfectly ordered classical state with wave vector $\vec Q$, the ordered
moment assumes its maximum value, $M(\vec Q)=S$, and we have
\begin{equation}
    \left\langle\vec S_{i}\vec S_{j}\right\rangle
    =
    \left\langle S_{i}^{z}S_{j}^{z}\right\rangle
    =
    S^{2}{\rm e}^{-{\rm i}\vec Q\left(\vec R_{i}-\vec R_{j}\right)},
    \quad i\ne j.
    \label{eqn:spins}
\end{equation}
Assuming perfect order for the finite tile under consideration, too,
we have
\begin{eqnarray}
    S_{N}(\vec Q)
    &=&
    \frac{N}{\cal N}
    \left[
    S(S+1)+
    (N-1)S^{2}\right]
    \nonumber\\
    &=&
    \frac{1}{\cal N}\ NS\left(NS+1\right).
\end{eqnarray}
If we require $S_{N}(\vec Q)=S^{2}$ also in this case, we have to 
set
\begin{equation}
    {\cal N}=N\left(N+\frac{1}{S}\right),
    \label{eqn:n}
\end{equation}
which is the normalization we use for any tile included into our
finite-size scaling analysis for $M_{N}(\vec Q)$.  This is in
accordance with Refs.~\onlinecite{schulz:96,bernu:94}, and slightly
deviates from the ${\cal N}=N^{2}$ normalization commonly used by many
authors.

In the FM regime, although the fully polarized (all-up) state is an
eigenstate of the Hamiltonian, the structure factor at the
antiferromagnetic ordering vectors remains small, but finite.
Assuming perfect order again, we get for the individual terms in the
sum Eq.~(\ref{eqn:sf})

\begin{equation}
    \left\langle\vec S_{1}\vec S_{1}\right\rangle=S(S+1),
    \quad
    \left\langle\vec S_{1}\vec S_{j}\right\rangle
    =
    S^{2},
    \quad
    j\ne1,
\end{equation}
leading to
\begin{equation}
    S_{N}(\vec Q)=\frac{N}{\cal N}
    \left[
    S(S+1)
    +
    S^{2}\sum_{j=2}^{N}{\rm e}^{-{\rm i}\vec Q\vec R_{j}}
    \right].
\end{equation}
Let us restrict to tiles with an even area $N$ which is a necessary
condition for being compatible with at least one of the non-FM phases
of the $J_{1a,b}$-$J_{2}$ model.  The sum in the above equation
contains only exponentials which can acquire the values $+1$ or $-1$
for $\vec Q=(\pi,\pi)$, $(\pi,0)$, or $(0,\pi)$.  For each of these
three wave vectors, there are $N/2$ sites $j$ with distance $\vec
R_{j}$ to site $1$ which have phase $+1$, and $N/2$ sites with phase
$-1$.  Site $1$ obviously belongs to the former, such that we have
$N/2-1$ terms left in the sum over the exponentials above having phase
$+1$, and we get
\begin{equation}
    \sum_{j=2}^{N}{\rm e}^{-{\rm i}\vec Q\vec R_{j}}
    =
    \left(\frac{N}{2}-1\right)\times(+1)
    +\frac{N}{2}\times(-1)
    =
    -1.
\end{equation}
With $\cal N$ given by Eq.~(\ref{eqn:n}), we therefore have
\begin{equation}
    S_{N}(\vec Q)
    =
    \frac{S}{N+1/S}
    \label{eq:snq}
\end{equation}
for all three antiferromagnetic wave vectors in the ferromagnetic
phase. As required, this value vanishes in the thermodynamic limit.

\subsection{Long-distance correlations}
\label{sect:longrange}

An alternative way of calculating the ordered moment in the
thermodynamic limit is mentioned for example in
Refs.~\onlinecite{bernu:94,sandvik:97}: For the infinite system, in an
ordered phase with a staggered moment, the spin correlation function
factorizes for large distances $|\vec R_{i}-\vec R_{j}|$, and
Eq.~(\ref{eqn:spins}) simplifies:
\begin{equation}
    \lim_{\left|\vec R_{i}-\vec R_{j}\right|\to\infty}
    \left|\left\langle\vec S_{i}\vec S_{j}\right\rangle\right|
    =\left|\left\langle\vec S_{i}\right\rangle
    \left\langle\vec S_{j}\right\rangle\right|
    =M^{2}(\vec Q).
\end{equation}
Working with a finite compact tile, we can extrapolate the spin
correlation function for a single pair of spins, defining lattice
points $i$ and $j$ such that
their distance on the tile is maximized under given periodic boundary
conditions.  Without loss of generality, we can restrict ourselves to
finding the pair $(1,j)$ or just site $j$ being maximally apart from
the origin.

We have to define a metric on a tile reflecting the periodic boundary
conditions: Each tile has four corners with coordinates $(0,0)$, $\vec
a_{1}$, $\vec a_{2}$, and $\vec a_{1}+\vec a_{2}$, which are all
equivalent to point $1$.  Using the Mannheim distance $d_{\text M}$
again, we define the {\em toroidal Mannheim distance\/} between two
points $\vec R_{i}$ and $\vec R_{j}$ on a tile as
\begin{widetext}
\begin{equation}
    d_{\text T}(i,j)=\min
    \left\{
    d_{\text M}\left(\vec R_{ij},0\right),
    d_{\text M}\left(\vec R_{ij},\vec a_{1}\right),
    d_{\text M}\left(\vec R_{ij},\vec a_{2}\right),
    d_{\text M}\left(\vec R_{ij},\vec a_{1}+\vec a_{2}\right)
    \right\}
\end{equation}
\end{widetext}
yielding the minimal distance between two points $\vec R_{i}$ and
$\vec R_{j}$ on a torus.  Here the vector $\vec R_{ij}$ is the point
emerging from $\vec R_{j}-\vec R_{i}$ when shifting $\vec R_{i}$ to
the origin, projected back into the original tile: With
\begin{eqnarray}
    \vec R_{j}-\vec R_{i}
    &=&
    \left(x_{j1}-x_{i1}\right)\vec a_{1}+
    \left(x_{j2}-x_{i2}\right)\vec a_{2}
    \nonumber\\
    &=&
    \frac{1}{N}
    \left[
    \left(n_{j1}-n_{i1}\right)\vec a_{1}+
    \left(n_{j2}-n_{i2}\right)\vec a_{2}
    \right]
\end{eqnarray}
this point is simply given by
\begin{eqnarray}
    \vec R_{ij}
    &=&
    \frac{1}{N}\left[
    \left(\left(n_{j1}-n_{i1}\right)\mathop{\rm mod}N\right)\vec a_{1}
    \right.
    \\
    &&\phantom{\frac{1}{N}}
    \left.
    +
    \left(\left(n_{j2}-n_{i2}\right)\mathop{\rm mod}N\right)\vec a_{2}
    \right].
    \nonumber
\end{eqnarray}
With respect to the distance $d_{\text T}$, we can then define
\begin{equation}
    \tilde M_{N}^{2}(\vec Q)
    =
    \left|\left\langle\vec S_{1}\vec S_{j}
    \right\rangle\right|_{j
    =
    \max_{d_{\text T}(1,j)}(j\in{\cal T}(N))}.
    \label{eq:Msecond}
\end{equation}
For a square with size $N=L^{2}$, $L$ even, the point $\vec
R_{j}=(L/2,L/2)$, i.\,e., the center of the square, has maximum
distance $d_{\text T}$ from the origin.  However, in most cases, due
to the lack of a lattice point located in the geometrical center of
the tile, there will be, for even $N$, {\em two\/} sites $j$ having
the same maximum distance $d_{\text T}$ from the origin, see
Fig.~\ref{fig:coords14} for an example.\footnote{For odd $N$, tiles
with squareness $\rho=1$ have {\em four\/} equivalent points with
maximum distance from the origin.} For our calculations, we just
select one of them.  We then can give an estimate for the ordered
moment as
\begin{equation}
    M^{2}(\vec Q)=
    \lim_{N\to\infty}
    \tilde M_{N}^{2}(\vec Q).
\end{equation}

\section{Finite-size scaling analysis}
\label{sect:fsscaling}

In Refs.~\onlinecite{hasenfratz:93,sandvik:97}, the area dependence of
ground-state properties of the two-dimensional antiferromagnetic
Heisenberg model has been derived using chiral perturbation theory. In
particular, for the ground-state energy and the ordered moment, the
following scaling behavior has been found:
\begin{eqnarray}
	E_{0N}&=&E_0+\beta c\frac{1}{N^{3/2}}+\frac{c^2}{4\rho}\frac{1}{N^{2}}+
	{\cal O}\left(\frac{1}{N^{5/2}}\right),
	\label{eqn:escl0}
    \\
    M_N^{2}(\vec Q)&=&
    M^{2}(\vec Q)+
    \alpha\frac{M^2(\vec Q)}{c\chi_\perp}\frac{1}{N^{1/2}}+
	{\cal O}\left(\frac{1}{N}\right),
	\label{eqn:mscl0}
\end{eqnarray}
where $c=\sqrt{\rho/\chi_\perp}$ is the spin-wave velocity, $\rho$ the spin
stiffness constant, and $\chi_\perp$ the transverse susceptibility.
$\alpha=0.620704$ and $\beta=-1.437745$ are numerical constants.

Although these results were proposed only for the isotropic unfrustrated case,
we use them for the whole phase diagram.  We argue that the form of the scaling
functions should not depend on range and anisotropy of interactions within
certain limits, as long as the model belongs to the same universality class.
However the individual coefficients in Eqs.~(\ref{eqn:escl0})
and~(\ref{eqn:mscl0}) might change. Hence we assume the following size
dependences for the ground-state energy and the ordered moment:
\begin{eqnarray}
    E_{0N}&=&E_0+\frac{e_1}{N^{3/2}}+\frac{e_2}{N^{2}},
    \label{eqn:escl}
    \\
    M_N^{2}(\vec Q)&=&
    M^{2}(\vec Q)+\frac{m_1^{2}}{N^{1/2}}+\frac{m_2^{2}}{N},
    \label{eqn:mscl}
\end{eqnarray}
where the latter scaling function is applied to both $M_{N}(\vec Q)$
and $\tilde M_{N}(\vec Q)$.  With the exception of the point
$(\phi,\theta)=(0,\pi/4)$ (isotropic nearest-neighbor exchange $J_{1}$
only), for all combinations of $\phi$ and $\theta$ discussed here we
use only the first two terms in the equations above for our scaling
analysis.  This is due to the comparatively small number of different
areas of the tiles actually included in the calculation.

We have calculated the ground-state energies $E_{0N}$, structure
factors $M_{N}^{2}(\vec Q)$, and long-distance correlation functions
$\tilde M_{N}^{2}(\vec Q)$ at the respective ordering vectors for tilings
of $11$ different sizes $N$ between $8$ and $28$ at more than $300$
points in the $(\phi,\theta)$ plane, producing roughly $17\,500$ data
sets altogether.  To demonstrate the procedure, in the following we
show a few examples at some significant points in the phase diagram
before presenting our main results in section~\ref{sect:results}.

\subsection{The ground-state energy}
\label{sect:energy}

The ground-state energy of the Hamiltonian~(\ref{eqn:ham}) is calculated in the
subspace with total spin $S_z=0$ at wave vector $\vec k=\vec Q$ (the respective
classical ordering vector) except for tiles with area $N=4\ell$,
$\ell\in\mathbb{N}$, where the ground state is found in the $\vec k=0$ sector.
Figs.~\ref{fig:energy-neel}--\ref{fig:energy-corner} show the dependence of the
ground-state energy per site $E_{0N}$ on the reciprocal area of the tiles for
selected model parameters $(\phi,\theta)$.  (We plot $-E_{0N}$ in order to be
consistent with previous studies~\cite{schulz:96,betts:99}.)  The insets display
schematically the position in the phase diagram the plots refer to. Here and in
the following, energies are measured in units of $J_{\text c}$, unless mentioned
otherwise.

The short horizontal dashes show the calculated energies for those tiles
compatible with the corresponding classical phases for a given parameter set
$(\phi,\theta)$. The energy eigenvalues of those tiles having maximum squareness
are indicated by the small open circles. We have also determined $E_{0N}$ for
tiles incompatible with the classical phases.  In these cases, we get
consistently much higher values for $E_{0N}$, and we omit them in the plots. 
The larger value of the energy for these tilings is due to the fact that the
interactions governed by the geometry of the tile prevent achieving a lower
value of the energy.

Furthermore there are tiles which are compatible with classical phases
but correspond to very skewed parallelograms with small squareness
(less than $1/2$), in some cases they are ladders or even chains.
They have a much lower ground-state energy (shown as dashes above the
scaling line in Figs.~\ref{fig:energy-neel} and~\ref{fig:energy-caf})
than the tiles of equal size and larger squareness.  Although the
energies of these finite clusters with low squareness are shown in the
figure for completeness they are not used for the scaling procedure.

For each tile area $N$, we have to choose one value for $E_{0N}$ out
of the stack of eigenvalues to be included into the finite-size
scaling fit applying Eq.~(\ref{eqn:escl}).  According to our
selection criteria, we choose the tile which (i) is compatible with
the classical phase the parameter set $(\phi,\theta)$ belongs to, and
(ii) is the best approximation to a square in the sense of
Eq.~(\ref{eqn:squareness}).  See Table~\ref{tbl:pg} for a list of them.
The values $E_{0N}$ obtained in this way, labeled by the open
circles in Figs.~\ref{fig:energy-neel}--\ref{fig:energy-corner},
are then used for a $\chi^{2}$ fit to Eq.~(\ref{eqn:escl}), indicated 
by the solid lines in the figures.

\subsubsection{Isotropic nearest-neighbor exchange}

\begin{figure}
    \includegraphics[width=\columnwidth]{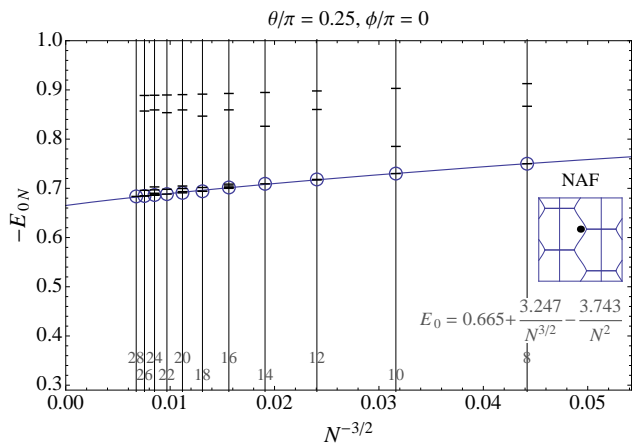}
    \includegraphics[width=\columnwidth]{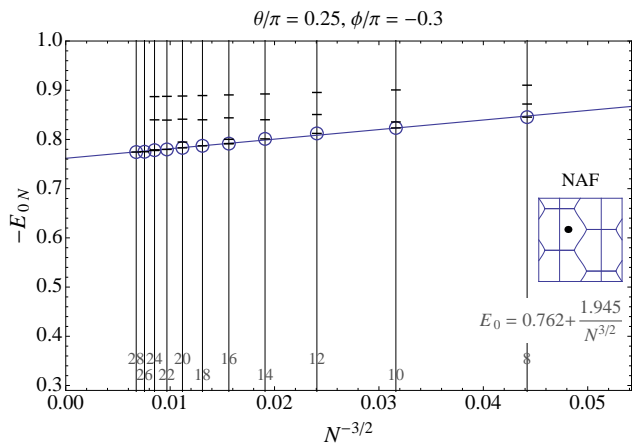}
    \caption{Ground-state energy of the isotropic model as a function
    of $1/N^{3/2}$.  Top: Conventional N\'{e}el AF Heisenberg model
    ($J_{1a}=J_{1b}$, $J_2=0$).  Bottom: Additional ferromagnetic
    next-nearest neighbor exchange $J_{2}<0$.  The dashes indicate the
    energy per site for different tilings.  Those having maximum
    squareness, marked with a circle, are used for the scaling and
    extrapolation.  The solid line shows the scaling curve,
    Eq.~(\ref{eqn:escl}).  The insets in these and the following plots
    show the parameter position $(\phi,\theta)$ in the phase diagram,
    cf.~Fig.~\ref{fig:phase}. The numbers near the abscissae denote 
    the tile size $N$.}
    \label{fig:energy-neel}
\end{figure}

The top of Fig.~\ref{fig:energy-neel} shows the ground-state energy scaling of
the conventional unfrustrated antiferromagnetic Heisenberg model with
$J_{1a}=J_{1b}=J_{\text c}$ and $J_2=0$, equivalent to
$(\phi,\theta)=(0,\pi/4)$.  We obtain the value $E_0=-0.66(5)$ for the
extrapolated ground state energy which is in agreement with previous
calculations, see, e.\,g., Refs.~\onlinecite{schulz:96,betts:99} and references
cited therein.  It should be noted that not always the tile having the highest
squareness also has the highest ground-state energy compared to the other tiles
with the same area $N$. However, the differences are small, and not visible on
the scale of Fig.~\ref{fig:energy-neel}.

\subsubsection{Next-nearest neighbors and spatial anisotropy}

The lower plot in Fig.~\ref{fig:energy-neel} shows the scaling of the
ground state energy for the point $(\phi,\theta)=(-0.3\pi,\pi/4)$
corresponding to the isotropic model with finite ferromagnetic
next-nearest-neighbor exchange $J_2$.  Ferromagnetic $J_{2}$
stabilizes the $(\pi,\pi)$ order, and we obtain a lower ground state
energy $E_0=-0.76(2)$ accordingly.

\begin{figure}
    \includegraphics[width=\columnwidth]{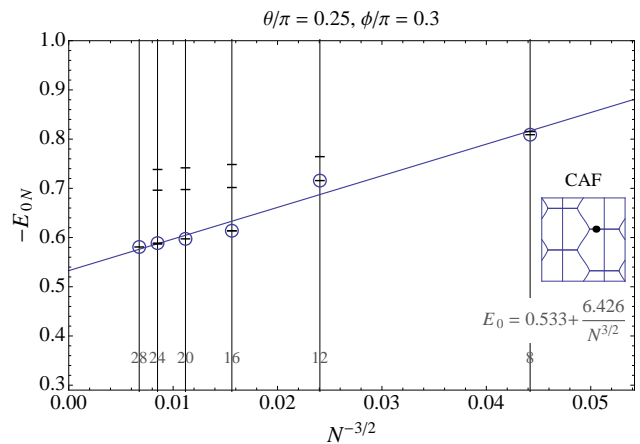}
    \includegraphics[width=\columnwidth]{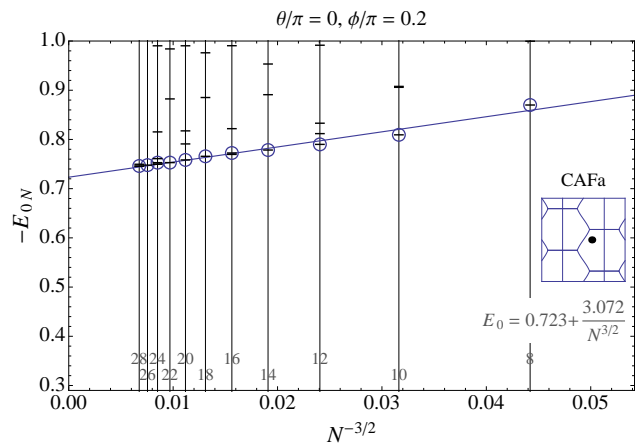}
    \caption{Finite-size scaling of the ground-state energy in
	(top) the CAF phase of the isotropic and (bottom) the CAFa phase of
	the maximally anisotropic model for different tilings. Labels, symbols and
	lines have the same meaning as in Fig.~\ref{fig:energy-neel}.}
    \label{fig:energy-caf}
\end{figure}

The top of Fig.~\ref{fig:energy-caf} illustrates the scaling behavior
in the CAF phase of the isotropic model with
$(\phi,\theta)=(0.3\pi,\pi/4)$.  Here all exchange constants are
antiferromagnetic, and the competition between them leads to columnar
order.  Because $J_{1a}=J_{1b}$ here, we have two equivalent ordering
vectors $(\pi,0)$ and $(0,\pi)$, labeled by CAFa and CAFb,
respectively.  Only the tiles which contain both the CAFa and CAFb
classical phases have truly the symmetries required by the Hamiltonian
in this case.  Consequently only the tiles with size $N=4\ell$,
$\ell\in\mathbb{N}$ are acceptable for scaling, and we have much less
data points available as compared to the Néel phase,
Fig.~\ref{fig:energy-neel}.  As before, we select out of these the tiles
with maximum squareness for the scaling procedure which leads to a
ground state energy $E_0=-0.53(2)$.

In the anisotropic case, CAFa and CAFb phases are no longer
equivalent.  An example of CAFa for the maximally anisotropic case
$(\phi,\theta)=(0.2\pi,0)$, where $J_{1b}=0$ and $J_{2}$
antiferromagnetic, is shown in the bottom part of
Fig.~\ref{fig:energy-caf}.  Here the tiles do not have to be
compatible with the $(0,\pi)$ ordering of the CAFb phase, therefore
the number of possible tilings is larger.  Also the scaling behavior
is improved again.  Compared to the isotropic case, the extrapolated
value for the ground-state energy is lower, indicating that the
introduction of a rectangular anisotropy stabilizes the columnar
order.

\subsubsection{Magnetically disordered regimes}

The scenarios discussed until here have one property in common: All of
them have parameter sets $(\phi,\theta)$ which are located deeply
inside the corresponding classically ordered phases, and all of them
show a good and well-defined scaling behavior of the ground-state
energy.  (We give a precise definition and discussion of the meaning
of ``good'' and ``well-defined'' in Sect.~\ref{sec:error}.)

\begin{figure}
    \includegraphics[width=\columnwidth]{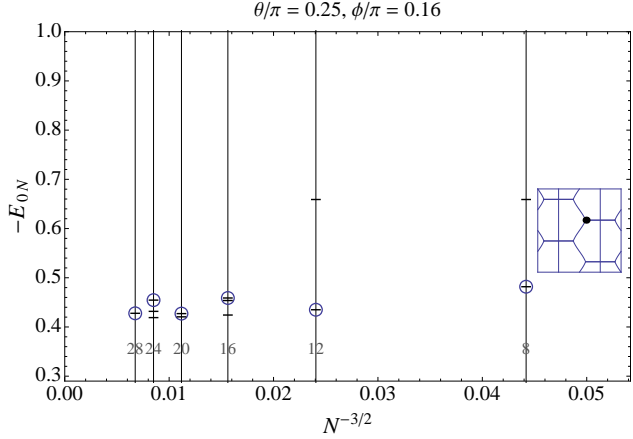}
    \caption{The tile size dependence of the ground-state energy in
    the disordered regime at the CAF-FM border.  In this case a
    reliable linear scaling result cannot be obtained from the
    circles.}
    \label{fig:energy-corner}
\end{figure}

However, this changes when approaching the disordered regimes at the borders of
the columnar phases.  Even with our stringent conditions for tile selection, the
behavior in the region where the transition from the N\'{e}el to the CAF phase
occurs is quite different.  In fact an area dependence as given by
Eq.~(\ref{eqn:escl}) no longer seems to apply, and the concept of choosing the
most-square-like tiles for scaling apparently becomes inappropriate.  An example
of this behavior is displayed in Fig.~\ref{fig:energy-corner} corresponding to
the well know disordered case of the isotropic model with $J_2/J_1\approx1/2$.

Expressed differently, finite-size scaling breaks down and is no longer a useful
concept by itself in and near the disordered regions at the edges of the
columnar phases in the phase diagram.  This will be described in more detail
including a discussion of the correlation functions and the ordered moment in
section~\ref{sect:results}.

We conclude that a stable finite-size scaling analysis of the ground state
energies can be achieved inside the NAF and  CAFa,b regions for all frustration
and anisotropy parameters provided a careful selection of tiles according to
phase compatibility and maximum squareness in each case is carried through.
However the scaling analysis cannot be made in the transition regions close to
the phase boundaries in Fig.~\ref{fig:phase}. This is correlated with the
breakdown of the ordered moment in these regions as will be shown in the next
section.

\subsection{Ordered moment, structure factor and correlation function}
\label{sect:structure}

Using the ground-state wave-function obtained by diagonalizing the
Hamiltonian matrix, we calculate the expectation values of the spin
correlation functions $\langle\vec S_i\vec S_j\rangle$.  Subsequently,
we calculate the finite-size equivalent $M_{N}^{2}(\vec Q)$ of the
ordered moment applying the two methods described in
Sects.~\ref{sect:structurefactor} and~\ref{sect:longrange}.  In the
same way as for the ground-state energy in the previous section, we
select the tile with maximum squareness for each tile area $N$ and fit
Eq.~(\ref{eqn:mscl}) to the resulting data points.  Again, apart from
the conventional Néel antiferromagnet we apply a linear scaling only,
ignoring the last term in Eq.~(\ref{eqn:mscl}).

\subsubsection{Isotropic nearest-neighbor exchange}

\begin{figure}
    \includegraphics[width=\columnwidth]{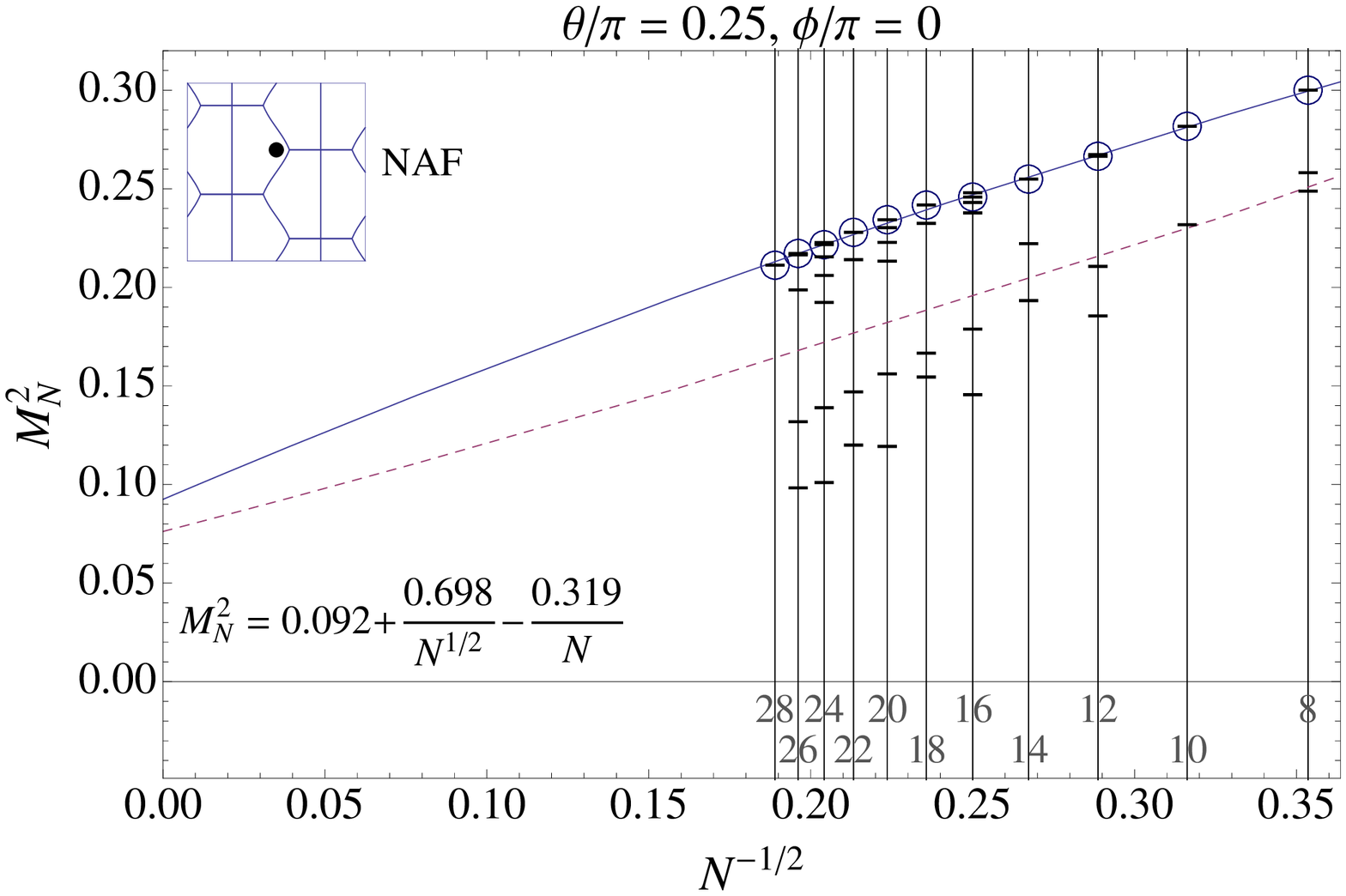}
    \includegraphics[width=\columnwidth]{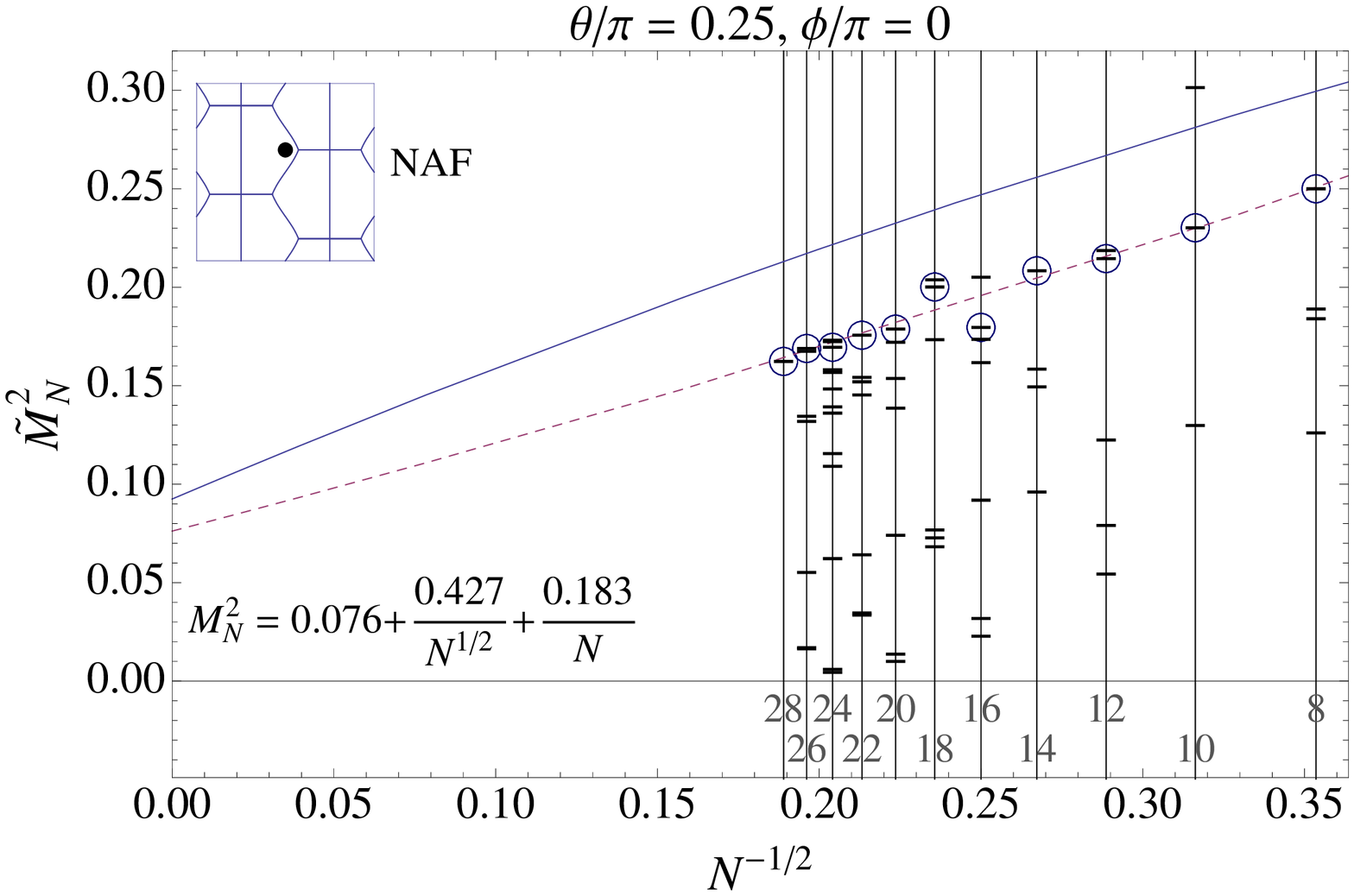}
    \caption{Finite-size scaling of the ordered moment for the
    isotropic nearest-neighbor AF Heisenberg model for different
    tilings.  Top: $M_{N}^{2}(\vec Q_{\text{NAF}})$ derived from the structure
    factor $S_{N}(\vec Q_{\text{NAF}})$.  Bottom: $\tilde M_{N}^{2}(\vec Q_{\text{NAF}})$
    derived from the long-distance correlation function.  The
    horizontal dashes and open circles denote the values for different
    tiles.  The latter are those with maximum squareness which are
    used for the scaling and extrapolation.  The solid lines show the
    fit of Eq.~(\ref{eqn:mscl}) to $M_{N}^{2}(\vec Q_{\text{NAF}})$, the dashed
    lines the fit to $\tilde M_{N}^{2}(\vec Q_{\text{NAF}})$. To facilitate the 
    comparison, we reproduce both fits in both plots.}
    \label{fig:moment-neel}
\end{figure}

The top part of Fig.~\ref{fig:moment-neel} shows the ordered moment derived from
the structure factor, Sect.~\ref{sect:structurefactor}, the bottom part shows
$\tilde M^{2}(\vec Q)$ obtained from the long-distance correlation function,
Sect.~\ref{sect:longrange}, for the unfrustrated ($J_2=0$) isotropic AF
Heisenberg model ($\phi=0,\theta=\pi/4$). As before, the horizontal dashes show
the values for different tilings, and the tiles with maximum squareness are
indicated by circles.  From the top figure it is evident that the most
square-like tiles have also the largest structure factor at the ordering vector
$\vec Q=(\pi,\pi)$ for the NAF phase.

A $\chi^{2}$ fit of Eq.~(\ref{eqn:mscl}) using all three terms has been applied
to both $M_{N}^{2}(\vec Q)$ and $\tilde M_{N}^{2}(\vec Q)$ separately.  The
solid lines in the two plots of Fig.~\ref{fig:moment-neel} denote the result of
fitting to $M_{N}^{2}(\vec Q)$, and we get $M(\vec Q_{\text{NAF}})=0.30(3)$.
This is in good agreement with previous
studies~\cite{schulz:96,sandvik:97,betts:99}.  The same fit applied to $\tilde
M_{N}^{2}(\vec Q)$, indicated by the dashed lines in the figure, gives a
slightly lower value $\tilde M(\vec Q_{\text{NAF}})=0.27(6)$ for the
thermodynamic limit.  We believe this to be a good indicator for the accuracy
within which we can determine asymptotic values for the ordered moment.  Also
the relative error (to be explained below) is larger for the latter method,
which is a consequence of the fact that only a single correlation function is
evaluated, while in the first method, a Fourier transform using all possible
$\langle\vec S_{i}\vec S_{j}\rangle$ is taken.

\subsubsection{Next-nearest neighbors and spatial anisotropy}

\begin{figure}
    \includegraphics[width=\columnwidth]{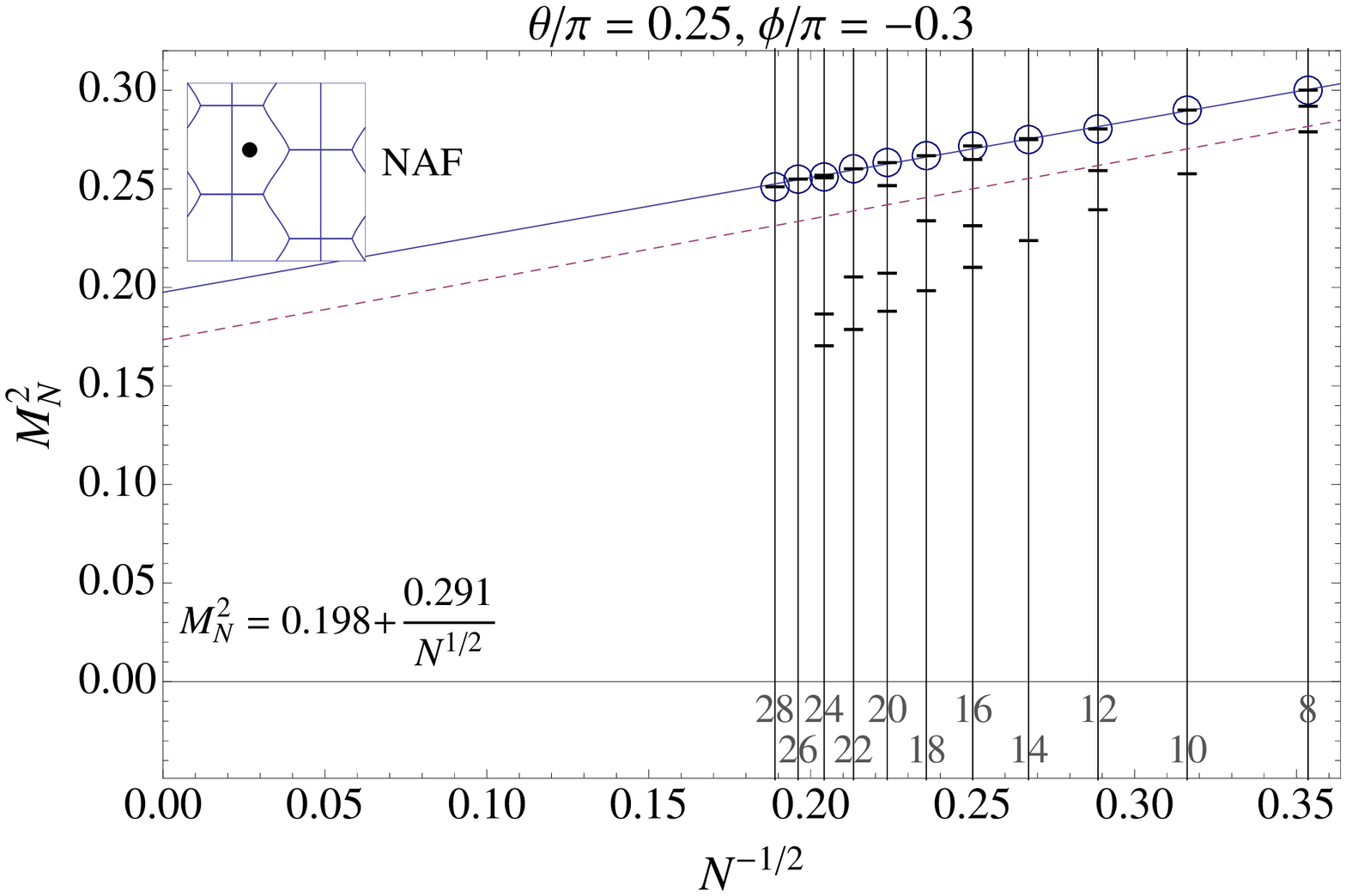}
    \includegraphics[width=\columnwidth]{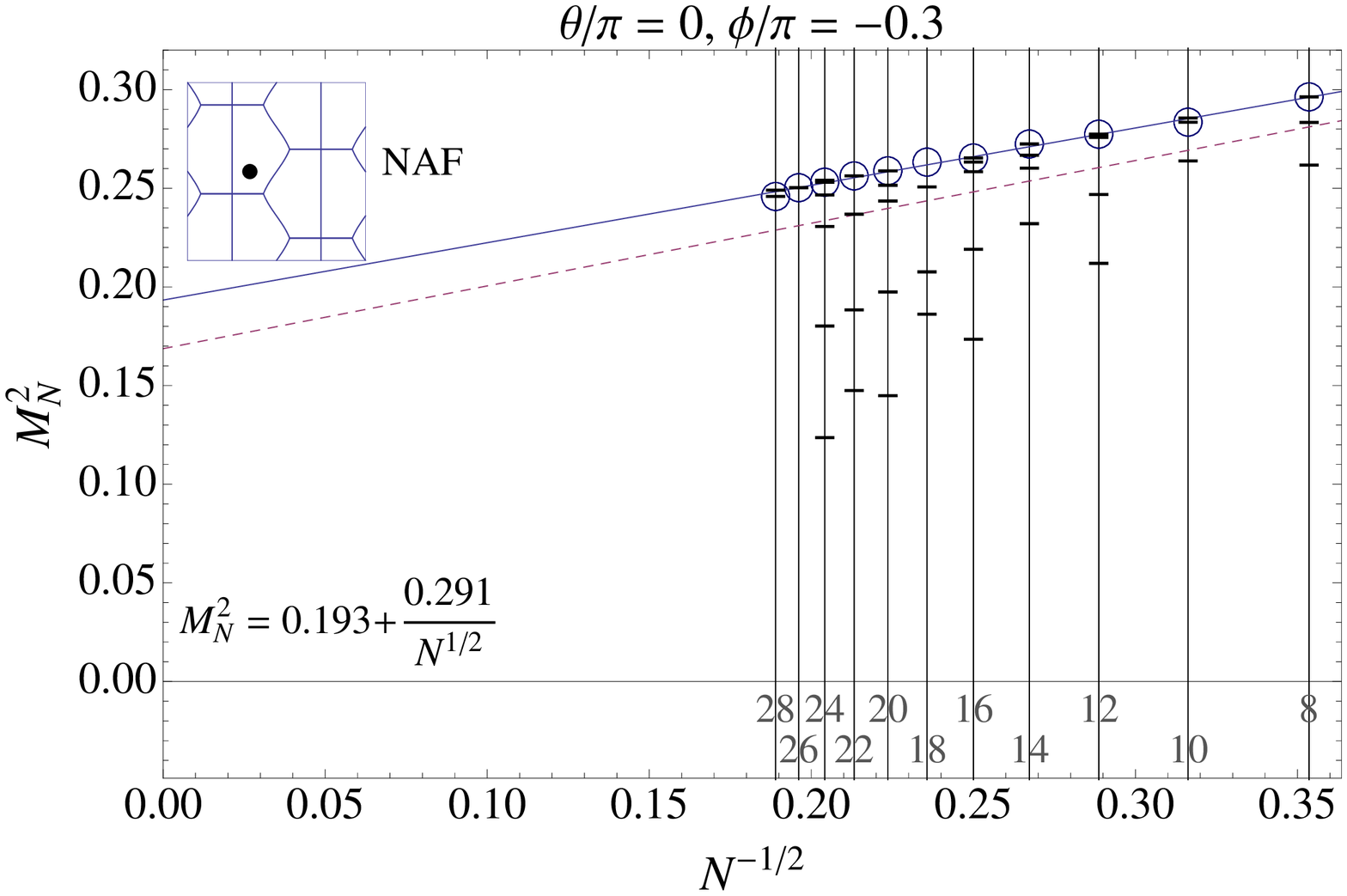}
    \caption{Finite-size scaling of the ordered moment in the NAF
    phase with ferromagnetic $J_{2}$.  Top: isotropic model,
    $J_{1a}=J_{1b}$.  Bottom: maximally anisotropic case, $J_{1b}=0$.
    The horizontal dashes denote $M_{N}^{2}(\vec Q)$ for different
    tilings, the solid lines are fits with Eq.~(\ref{eqn:mscl}) to
    the values for most-square-like-tiles, marked with a circle.
    Results for $\tilde M_{N}^{2}(\vec Q)$ for individual tiles are
    not shown, fits to $\tilde M_{N}^{2}(\vec Q)$ are displayed as the
    dashed lines.}
    \label{fig:moment-naf}
\end{figure}

The ordered moment scaling for ferromagnetic $J_2<0$ ($\phi=-0.3\pi$), both in
the isotropic ($J_{1a}=J_{1b}$) and maximally anisotropic ($J_{1b}=0$) case is
shown in the top and bottom of Fig.~\ref{fig:moment-naf} respectively. The
horizontal dashes denote $M_{N}^{2}(\vec Q)$ for individual tilings, the solid
line represents a fit with Eq.~(\ref{eqn:mscl}) to the values for
most-square-like-tiles (circles in the plot).  The dashed lines denote fits to
$\tilde M_{N}^{2}(\vec Q)$ for the same set of tiles. (Individual values are not
shown.)  Again, a comparison of the extrapolated values for $M^{2}(\vec Q)$ from
the two different scaling procedures can serve as an indicator of the quality of
the finite-size scaling analysis.

\begin{figure}
    \includegraphics[width=\columnwidth]{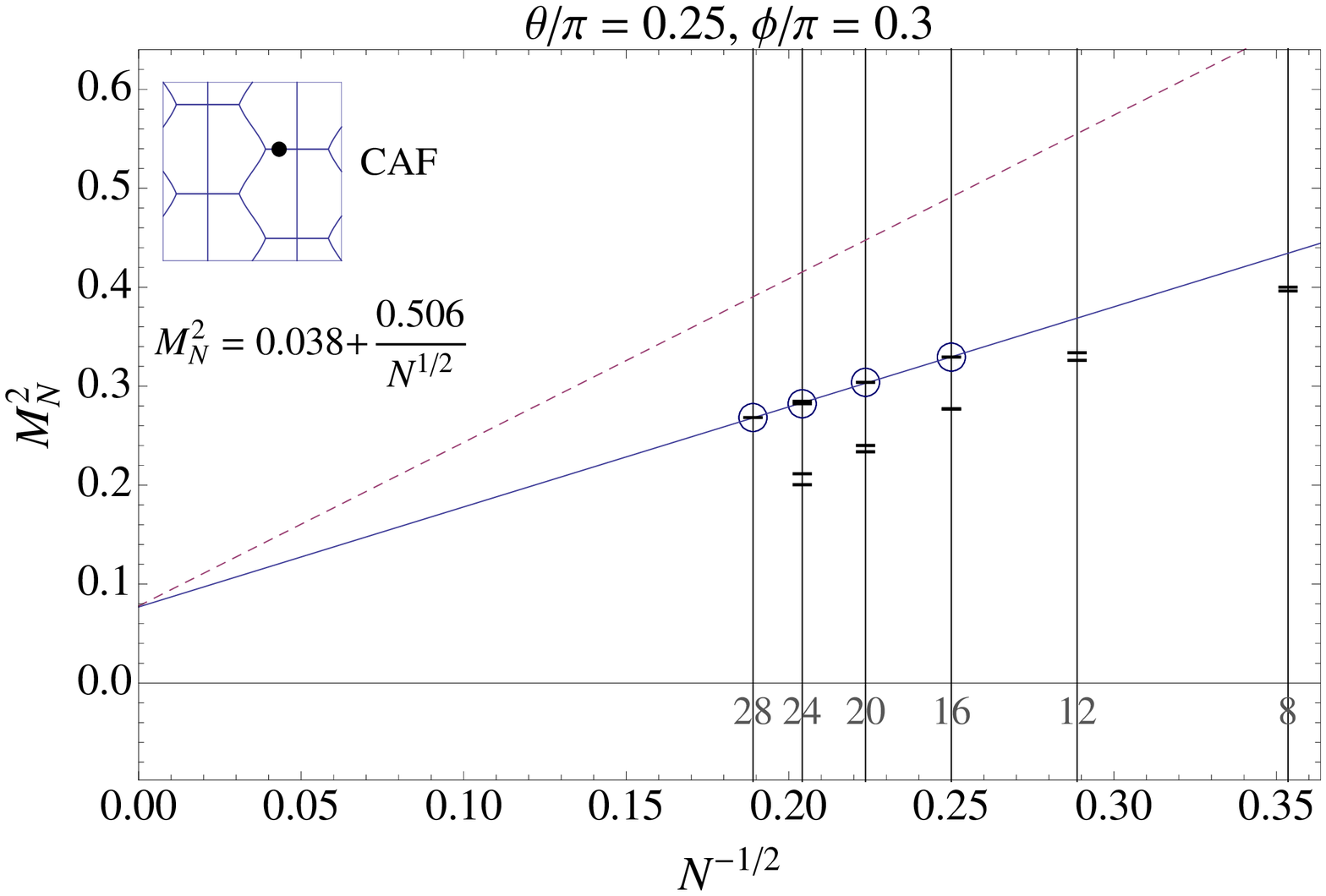}
    \includegraphics[width=\columnwidth]{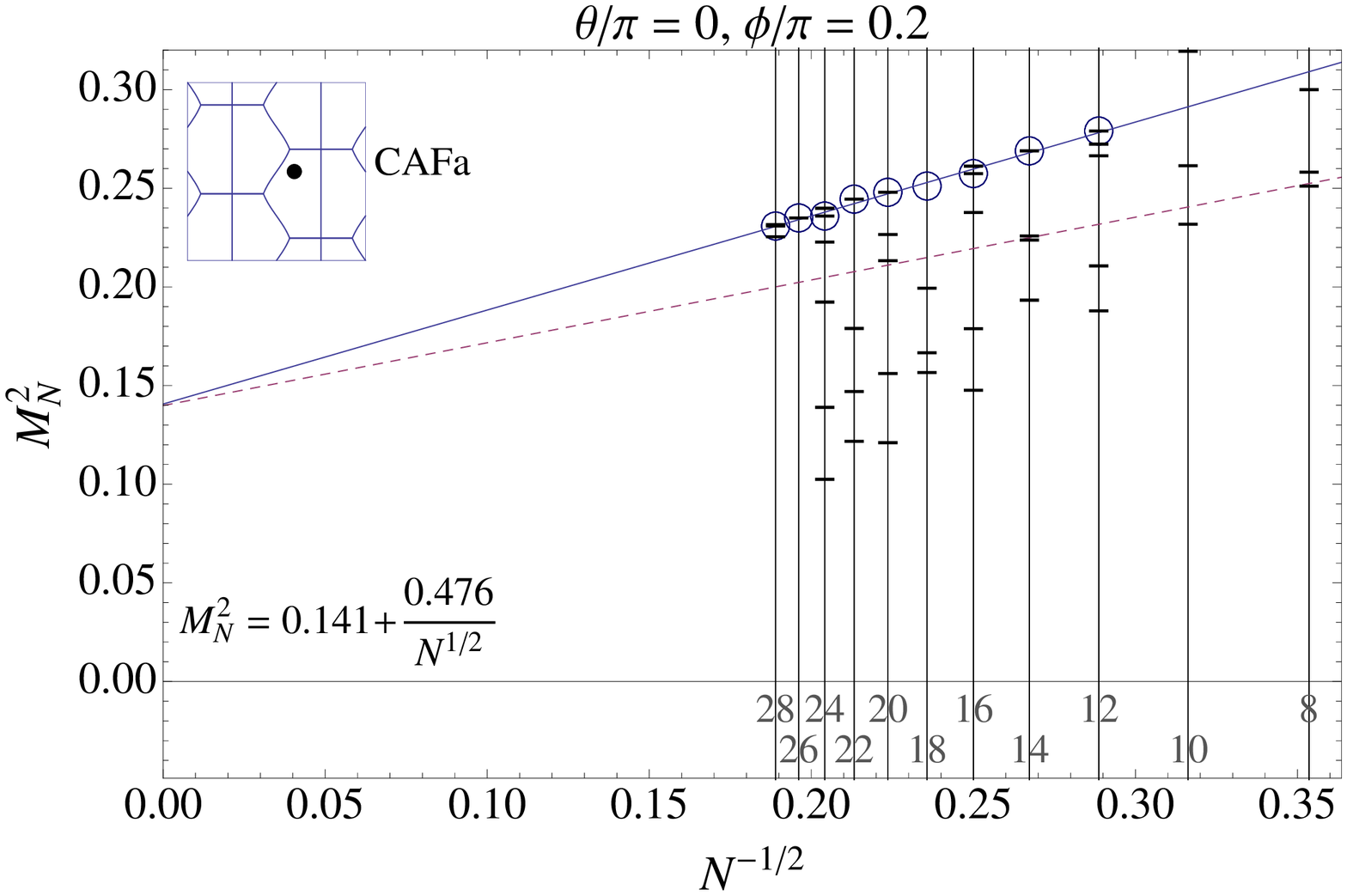}
    \caption{Finite-size scaling of the ordered moment in the columnar
    phase with antiferromagnetic $J_{2}$.  Top: isotropic model,
    $J_{1a}=J_{1b}$.  Bottom: maximally anisotropic case, $J_{1b}=0$.
    The horizontal dashes denote $M_{N}^{2}(\vec Q)$ for different
    tilings, the solid lines are fits with Eq.~(\ref{eqn:mscl}) to the
    values for most-square-like-tiles, marked with a circle.  Results
    for $\tilde M_{N}^{2}(\vec Q)$ for individual tiles are not shown,
    fits to $\tilde M_{N}^{2}(\vec Q)$ are displayed as the dashed
    lines.  The extrapolated values obtained from the scaling
    correspond to half the value of the ordered moment, due to the
    spatial symmetry breaking introduced with columnar ordering.}
    \label{fig:moment-caf}
\end{figure}

Fig.~\ref{fig:moment-caf} shows the scaling of the ordered moment in the
columnar phase for antiferromagnetic $J_{2}$.  For the isotropic case (top), as
before only those tiles compatible with both columnar phases, equivalent to
compatibility with all four classical phases can be used.  In the infinite
system, which has ${\cal C}_{4v}$ point-group symmetry, the wave vectors
$(\pi,0)$ and $(0,\pi)$ are equivalent.  However, most finite tiles have a
spatial symmetry lower than ${\cal C}_{4}$ (see Table~\ref{tbl:pg}), meaning
that the equivalence between $(\pi,0)$ and $(0,\pi)$ is lost.  For this reason
we take the sum of the structure factor at these two wave vectors and use the
resulting data points for scaling.  This is not necessary for the anisotropic
case ($\theta\ne\pi/4,-3\pi/4$), of which an example is shown at the bottom of
Fig.~\ref{fig:moment-caf}, since then CAFa and CAFb are different phases anyway.
 Here also a larger number of clusters is available for the scaling. Note that
in Fig.~\ref{fig:moment-caf} we plot the area dependence of the structure factor
and the long-distance correlation function including the factor $\zeta(\vec
Q)=2$ introduced in Eq.~(\ref{eq:Mfirst}), which is due to the spatial symmetry
breaking induced by the columnar order.

\subsubsection{Magnetically disordered regimes}

\begin{figure}
    \includegraphics[width=\columnwidth]{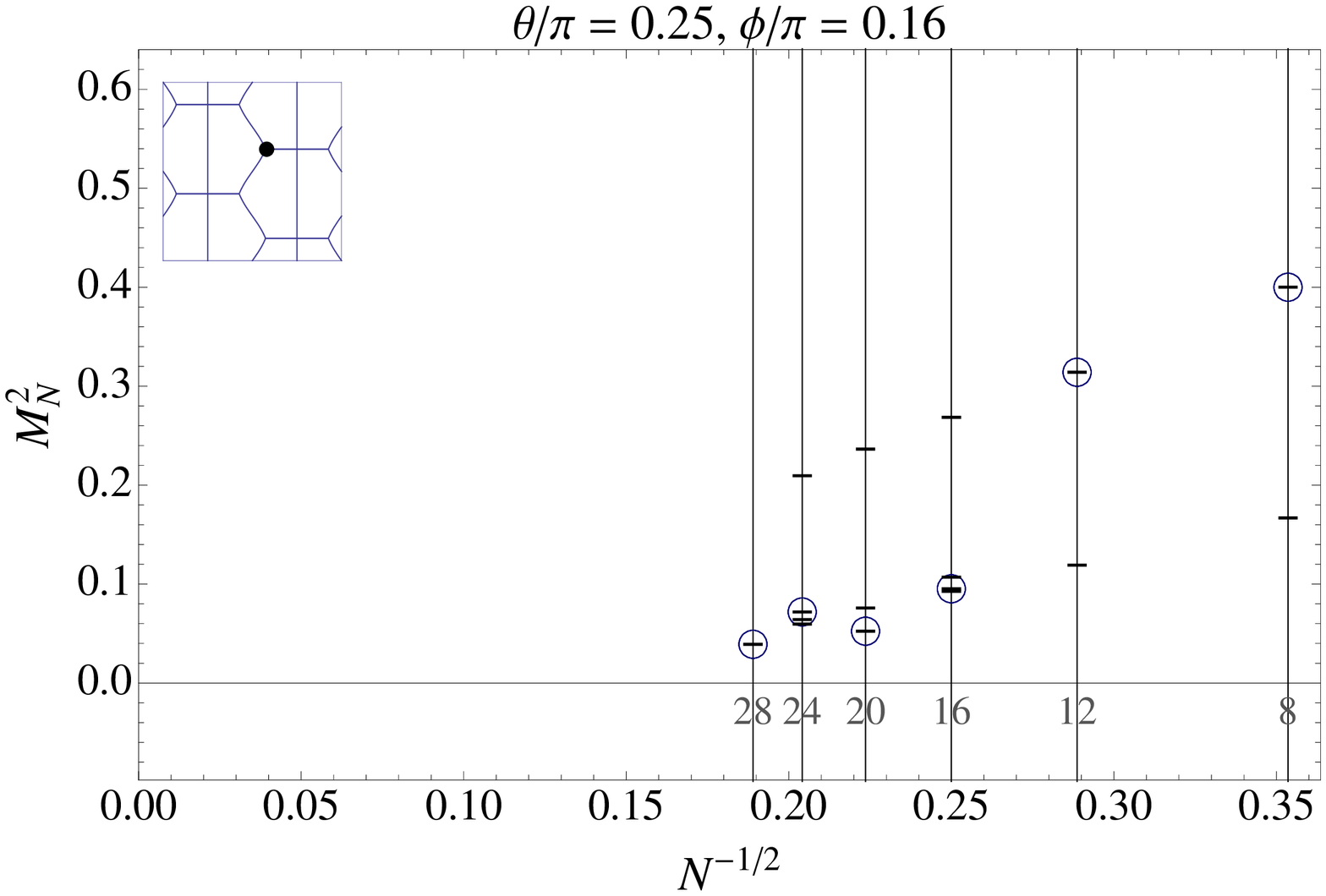}
    \includegraphics[width=\columnwidth]{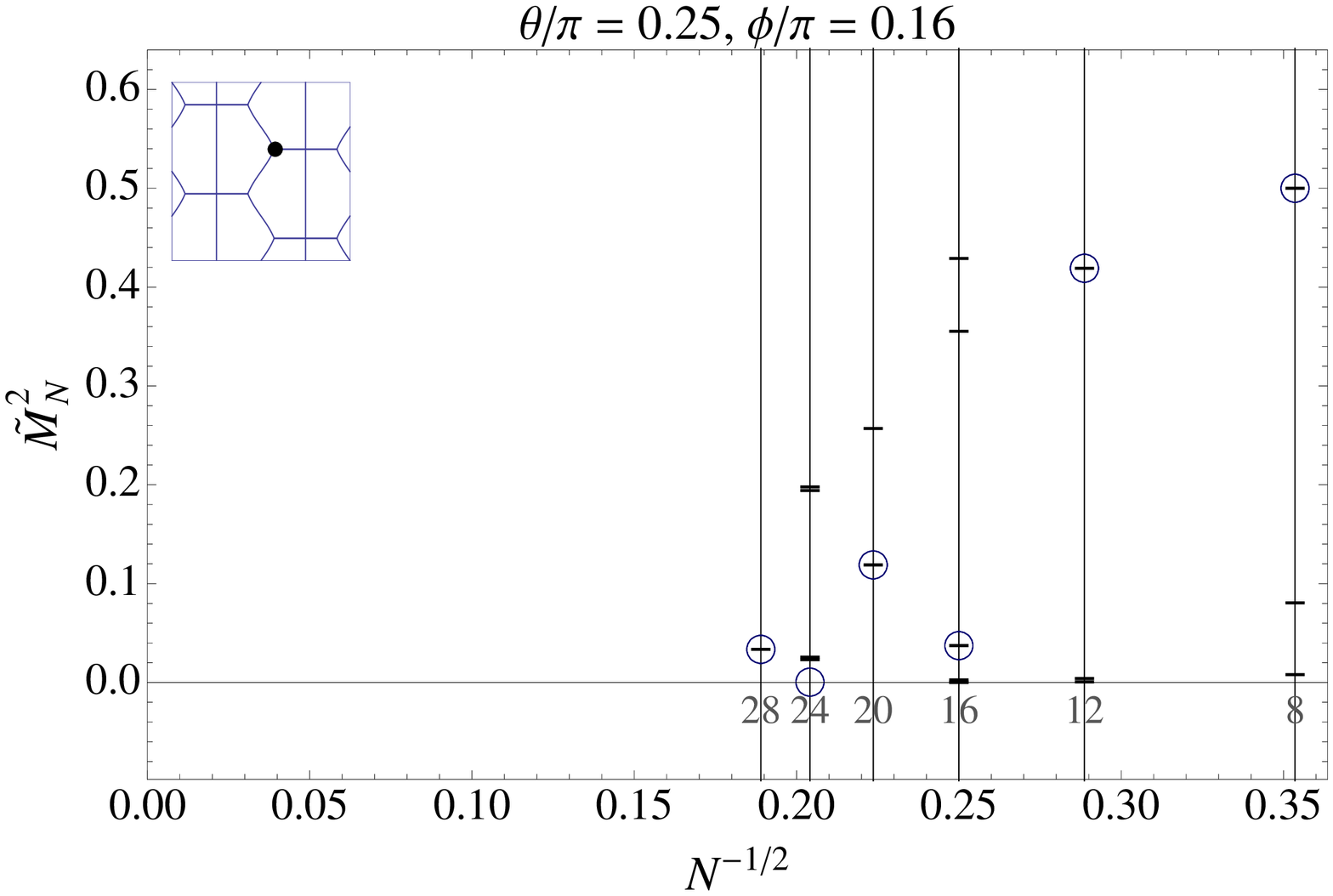}
    \caption{Tile area dependence of the ordered moment in the
    disordered regime at the NAF-CAF border for isotropic
    nearest-neighbor exchange.  The values are obtained from (top) the
    structure factor and (bottom) the long-distance correlation
    function, and the circles show the values for the tiles with
    maximum squareness.  No finite-size scaling is possible.}
    \label{fig:moment-corner}
\end{figure}

As it has been the case for the ground-state energy, in the disordered region
between the N\'{e}el and the columnar phase of the isotropic model
($J_{1a}=J_{1b}=J_{1}$, $J_{2}/J_{1}\approx1/2$), the behavior of both the
structure factor and the correlation function do not show a systematic
dependence on the tile size.  This is obvious from Fig.~\ref{fig:moment-corner},
which shows the calculated values for both $M_{N}^{2}(\vec Q)$ (top part) and
$\tilde M_{N}^{2}(\vec Q)$ (bottom part).  Disregarding for once the squareness
of our tiles, it appears that there are two ``stripes'' of values as a function
of $1/\sqrt{N}$ which both seem to extrapolate roughly to $0$.  However, a
quantitative analysis in the sense of Eq.~(\ref{eqn:mscl}) cannot be made.

\begin{figure}
    \includegraphics[width=\columnwidth]{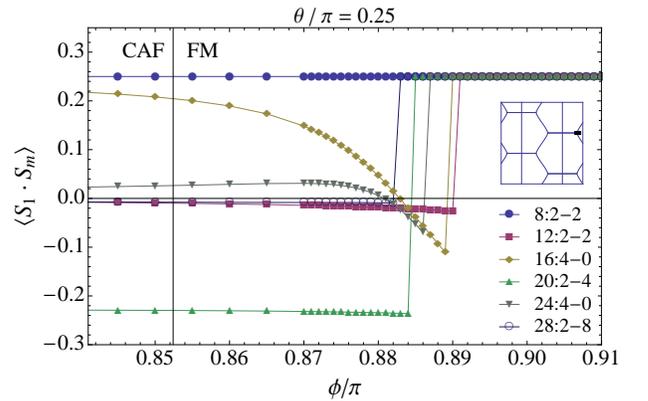}
    \caption{Dependence of the long-distance correlation function on
    the frustration angle $\phi$ in the spin nematic region between
    the CAF and FM phases of the isotropic model for those tiles with
    maximum squareness which are compatible to all four classical
    phases.  The legend in the plot lists the individual tiles and
    their symbols.}
    \label{fig:correl-iso-caf2fm}
\end{figure}

We observe a very similar behavior also for $J_{2}/J_{1}\approx-1/2$ (with
ferromagnetic $J_{1}$) in the spin-nematic region of the phase diagram.  In
order to see more details, the long-distance correlation function $\langle\vec
S_{1}\vec S_{m}\rangle$, used in Eq.~(\ref{eq:Msecond}), in this narrow region
between CAF and FM phase is plotted as a function of the frustration angle
$\phi$ in Fig.~\ref{fig:correl-iso-caf2fm} for tiles of different size.  The
labels of the individual tiles are given in the legend of the plot. The solid
vertical line indicates the classical phase boundary.  Apart from the eight-site
tile, which is too small anyway because sites $1$ and $m$ are just two hops
apart, for each tile $\langle\vec S_{1}\vec S_{m}\rangle$ is monotonously
decreasing before jumping to the ferromagnetic value $\langle\vec S_{1}\vec
S_{m}\rangle=S^{2}=1/4$. Those correlation functions which are ferromagnetic
(positive, both sites $1$ and $m$ on the same sublattice) in the columnar phase
even change their sign before the jump.  This sign change possibly can serve as
an indicator of the breakdown of columnar order.  However, the transition to
ferromagnetic correlations in the region of $0.881\pi<\phi<0.891\pi$ seen here
is not happening uniformly for all tiles, hence it cannot be used to extrapolate
to the thermodynamic limit.  This will be discussed more quantitatively in the
next section.

We can draw the same conclusions for the ordered moment scaling as we
did before for the ground state energy: If we carefully select the
allowed clusters with appropriate classical phase compatibility and
maximum squareness we can carry out a well defined scaling procedure
to the thermodynamic limit for the stable NAF and CAFa,b regions.
However in the disordered regions at the borders of the CAFa/b
honeycomb (see Fig.~\ref{fig:phase}) scaling is impossible, which
is an indication of the imminent breakdown of magnetic order due to
quantum fluctuations. Similar conclusions were obtained already from 
linear spin wave theory~\cite{schmidt:10}.

\subsection{Accuracy of extrapolated values}
\label{sec:error}

For the linear scaling (Eqs.~(\ref{eqn:escl}) and~(\ref{eqn:mscl})
without the last term) applied to extrapolating the results to the
infinite lattice, an analysis of the quality of the scaling is
essential.  There are several measures that can be introduced to
characterize the quality of the scaling.  A convenient measure is the
relative error in the $\infty$-norm, generally defined as
\begin{eqnarray}
    \epsilon_{\text{rel}}
    &=&
    \frac{{\|f-d\|}_{\infty}}{{\|f\|}_{\infty}}
    \approx 10^{-p},
    \label{eqn:error}
    \\
    \|\{x_{1},x_{2},\dots\}\|_{\infty}
    &=&
    \max\left(|x_{1}|,|x_{2}|,\dots\right).
    \nonumber
\end{eqnarray}
Here, $f$ is the set of fitted values at the points $1/N$, and $d$ the
set of data calculated for the maximum-squareness tiles with area $N$.
The exponent $p$ in the above equation can be interpreted as the
number of significant digits~\cite{golub:96} for the extrapolated
value $f_{0}$ at $N\to\infty$.

\begin{figure}
    \includegraphics[width=\columnwidth]{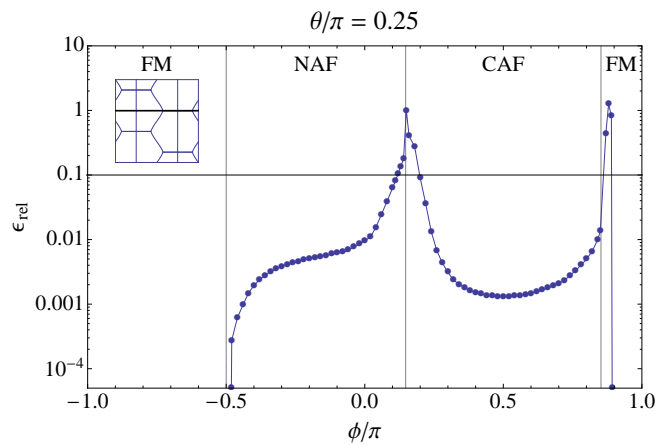}
    \caption{Relative error $\epsilon_{\text{rel}}$ as defined in
    Eq.~(\ref{eqn:error}) of the fits to $M_{N}^{2}(\vec Q)$ for the
    isotropic model.  Note the logarithmic scale of the ordinate.  The
    error in the magnetically disordered regions is at least one order of
    magnitude larger than in the ordered sectors of the phase diagram. The
    solid horizontal line denotes the value $\epsilon_{\text{rel}}=0.1$,
    indicating the maximum error acceptable for having at least one significant
    digit in $M^2(\vec Q)$.}
    \label{fig:error-theta-25}
\end{figure}

Figure~\ref{fig:error-theta-25} shows a plot of $\epsilon_{\text{rel}}$ for
$M_{N}^{2}(\vec Q)$ of the isotropic model as function of the frustration angle
$\phi$.  In the ferromagnetic phase, the fully polarized state is the
ground-state and the error reflects the numerical noise, i.\,e., the accuracy of
the floating point operations, which are of order $10^{-15}$, and has been
excluded from the figure.  (We have determined $M_{N}^{2}(\vec Q)$ for
$(\phi,\theta)$ values in the ferromagnetic region mainly in order to verify the
correctness of our numerical implementation of Eq.~(\ref{eqn:mscl}).)

In the well-ordered region of NAF and CAF we have $\epsilon_{\text{rel}}={\cal
O}(10^{-3}\dots10^{-2})$, and we can regard the scaling procedure in these areas
as stable.  This is in contrast to the magnetically disordered regions at the
borders of the columnar phases, where the strong increase in
$\epsilon_{\text{rel}}$ clearly indicates that the scatter of the points is much
higher and makes the extrapolated value unreliable. The solid horizonal line in
Fig.~\ref{fig:error-theta-25} denotes $\epsilon_{\text{rel}}=0.1$. According to
Eq.~(\ref{eqn:error}), this is the maximum error at which the extrapolated
values for $M^2(\vec Q)$ have at least one significant digit. In other words, a
magnetic order parameter $M^2(\vec Q)$ to be used to characterize the nature of
the ground state cannot be obtained anymore in those regions with
$\epsilon_{\text{rel}}>0.1$.

\section{Results}
\label{sect:results}

Here we describe the systematic dependence of the thermodynamic limit
of the ground state energy $E_{0}$ and the ordered moment $M_{0}(\vec
Q)$ on the model parameters $(\phi,\theta)$ as obtained from our
finite-size analysis discussed in the previous section.

\subsection{Ground-state energy}
\label{sect:gsenergy}

\begin{figure}
    \includegraphics[width=\columnwidth]{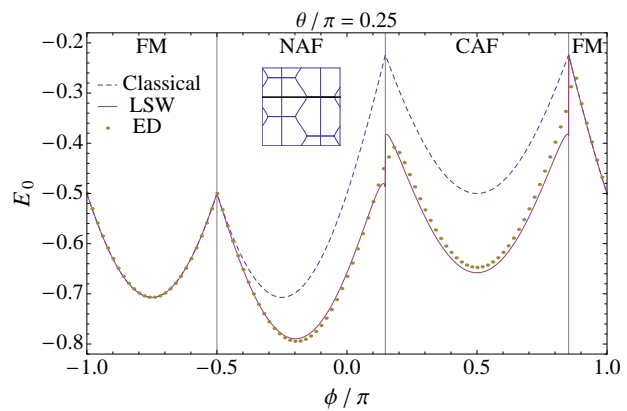}
    \caption{The ground-state energy as function of the frustration
    angle $\phi$ for the isotropic model with fixed $\theta=\pi/4$ .
    The classical energy is shown as dashed line, and the spin-wave
    results including zero-point fluctuations are presented as solid
    line.  Dots indicate the values for the energy $E_{0}$ obtained
    from extrapolating our exact-diagonalization data.}
    \label{fig:energy-theta-25}
\end{figure}

Fig.~\ref{fig:energy-theta-25} shows the energy dependence on the frustration
angle $\phi$ for the isotropic model with $\theta/\pi=1/4$.  The dashed line
shows the classical energy, the solid line displays the result from linear spin
wave theory which includes corrections due to zero-point fluctuations of
magnons~\cite{schmidt:10}.  The dots denote our scaled ED results
$E_{0}=E_{0}(\phi,\theta)$ according to Eq.~(\ref{eqn:escl}).

Inside the magnetic phases, the overall agreement between the numerical data and
the results obtained from linear spin-wave calculations~\cite{schmidt:10} (solid
line) is surprisingly good, and improves on the comparison with exact
diagonalization results from just a single cluster we have made
previously~\cite{schmidt:10}. However, in the disordered regions at the borders
of the columnar phase, which is shaded with gray, the fit to the numerical data
has a comparatively poor quality, and the reliability of the numerical result is
somewhat questionable. Linear spin-wave theory breaks down here, too, albeit in
a slightly different parameter range~\cite{schmidt:10}.

\begin{figure}
    \includegraphics[width=\columnwidth]{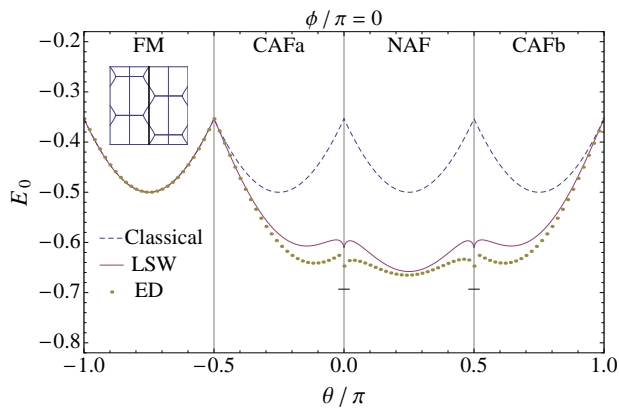}
    \caption{Ground state energy for
    the anisotropic model with first neighbor interaction only ($J_2=0$), as
    function of the anisotropy parameter $\theta$. The dashed line represents
    the energy of the classical model, the solid line includes corrections from
    linear spin wave theory. The extrapolated values from finite-size scaling
    of the ED results are shown as dots. Dashes denote the result $E_0=J_{\text
    c}\mathop{\rm ln}2$ from Bethe ansatz calculations for the one-dimensional $S=1/2$
    spin chain.}
    \label{fig:energy-phi-0}
\end{figure}

Next we turn to the unfrustrated ($J_2=0$) but anisotropic ($J_{1a}\ne J_{1b}$)
model with only next-nearest neighbor interactions. Fig.~\ref{fig:energy-phi-0}
shows the dependence of the ground-state energy per site $E_0$ on the anisotropy
parameter $\theta$ for fixed frustration angle $\phi=0$. The classical
ground-state energy is shown as a dashed line for comparison. As before, the
agreement with linear spin-wave theory (solid line) inside the magnetic phases
is good, which is not the case at the borders of the Néel phase.

But in this case, this has a reason different from the frustration induced by
competing interactions discussed before: For $\theta=0$ or $\theta=\pi/2$, the
nearest-neighbor exchange along one particular spatial direction vanishes,
$J_{1a}$ or $J_{1b}$, respectively, such that we are actually dealing with an
array of decoupled spin chains instead of a two-dimensional system. The
classical approximation fails to describe the ground-state of the chains
completely, and LSW corrections to it due to zero-point fluctuations are largest
at these two points, improving the classical result drastically. The
extrapolated ground-state energy from ED gives an even better approximation to
the true value for $E_0=J_{\text c}\mathop{\rm ln}2$ derived from Bethe ansatz
results~\cite{hulthen:38}, displayed as short horizontal dashes in
Fig.~\ref{fig:energy-phi-0}, but also here the agreement is limited. However
this is an extreme case, and we discuss it here primarily to show the limits of
the finite-size scaling method when applied to the strongly anisotropic model.

\subsection{Structure factor and ordered moment}
\label{sect:moment}

\begin{figure}
    \includegraphics[width=\columnwidth]{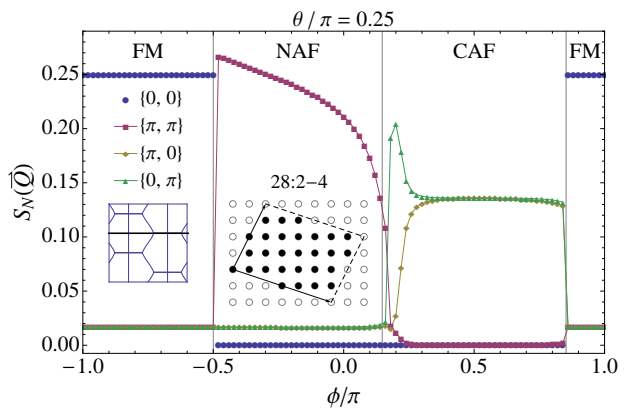}
    \caption{Structure factor $S_N(\vec Q)$ for tile 28:2-4 (inset:
    illustration of the compact tile) at the wave vectors corresponding to the
    four classically ordered phases as a function of the frustration angle $\phi$
    for the isotropic model. The labeling is explained
    in the inset of the figure.}
    \label{fig:sq-28:2-4}
\end{figure}

The calculated values for the structure factor for all four ordering wave
vectors as obtained using the Eq.~(\ref{eqn:sf}), for the largest system we are
considering (tile 28:2-4 with 28 spins) is shown in Fig.~\ref{fig:sq-28:2-4}.
Starting from the FM phase with a fully polarized ground state, we have
$S_N(\vec Q_{\text{FM}})=S^2$, and at the antiferromagnetic wave vectors $\vec
Q$ we get $S_N(\vec Q)=1/60$ in accordance with Eq.~(\ref{eq:snq}).

Since the shape of tile 28:2-4 is a parallelogram, and hence has ${\cal C}_2$
point-group symmetry only, the equivalence between the two wave vectors $\vec
Q=(\pi,0)$ and $(0,\pi)$ present for the infinite system (or any tile having at
least ${\cal C}_4$ point-group symmetry) is lost. This manifests itself in a
different $\phi$ dependence of the two structure factors $S(\vec Q_{\text{CAFa}})\ne
S(\vec Q_{\text{CAFb}})$ in the columnar phase, see Fig.~\ref{fig:sq-28:2-4}.
 
Moreover, the discontinuity between the value of the structure factor at the
$(0,0)$ ordering in the FM phase and the $(\pi,\pi)$ ordering in the NAF phase
($\phi \approx -\pi/2$) is a finite-size effect and will be suppressed by
increasing the cluster size.

\begin{figure}
    \includegraphics[width=\columnwidth]{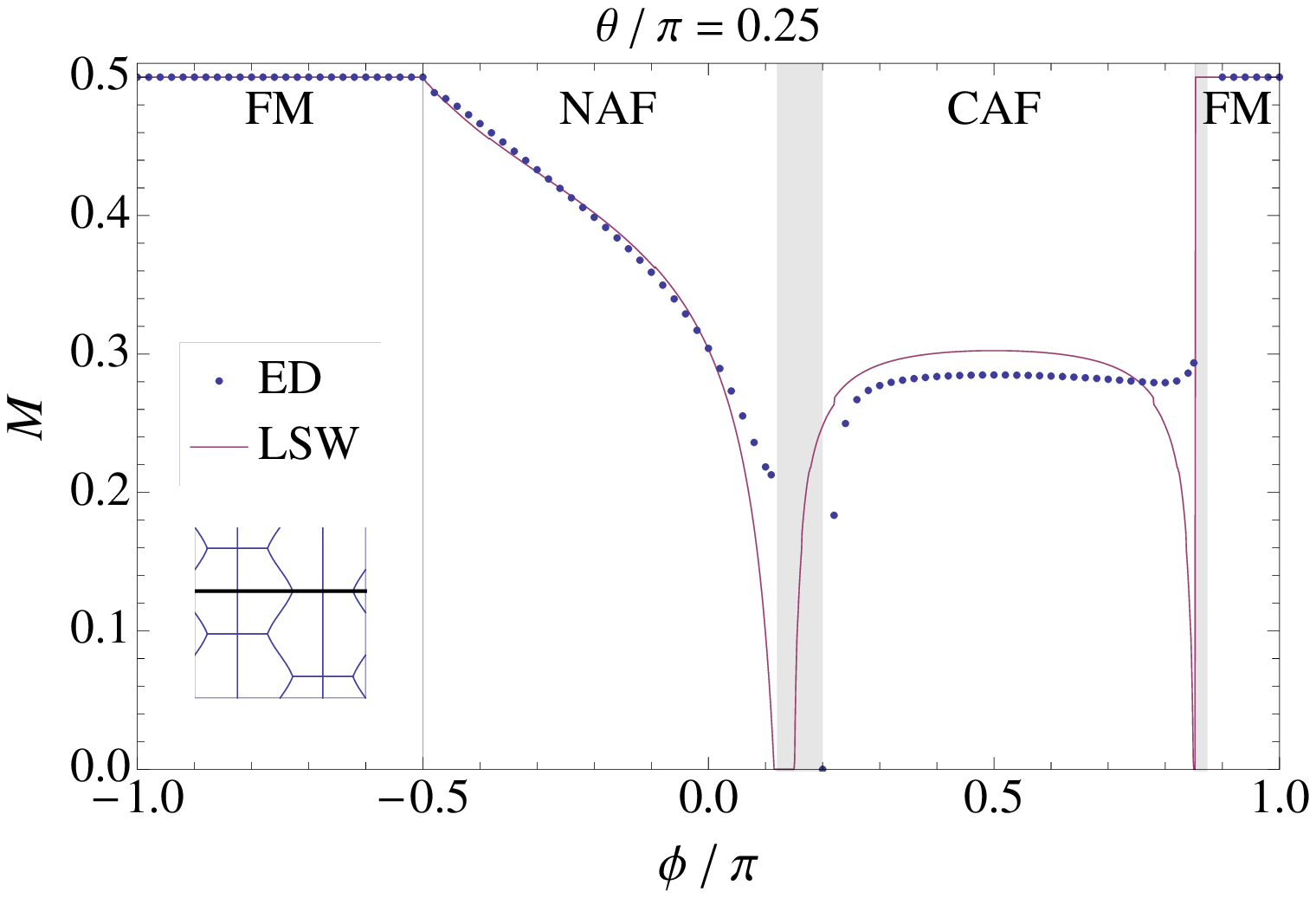}
    \includegraphics[width=\columnwidth]{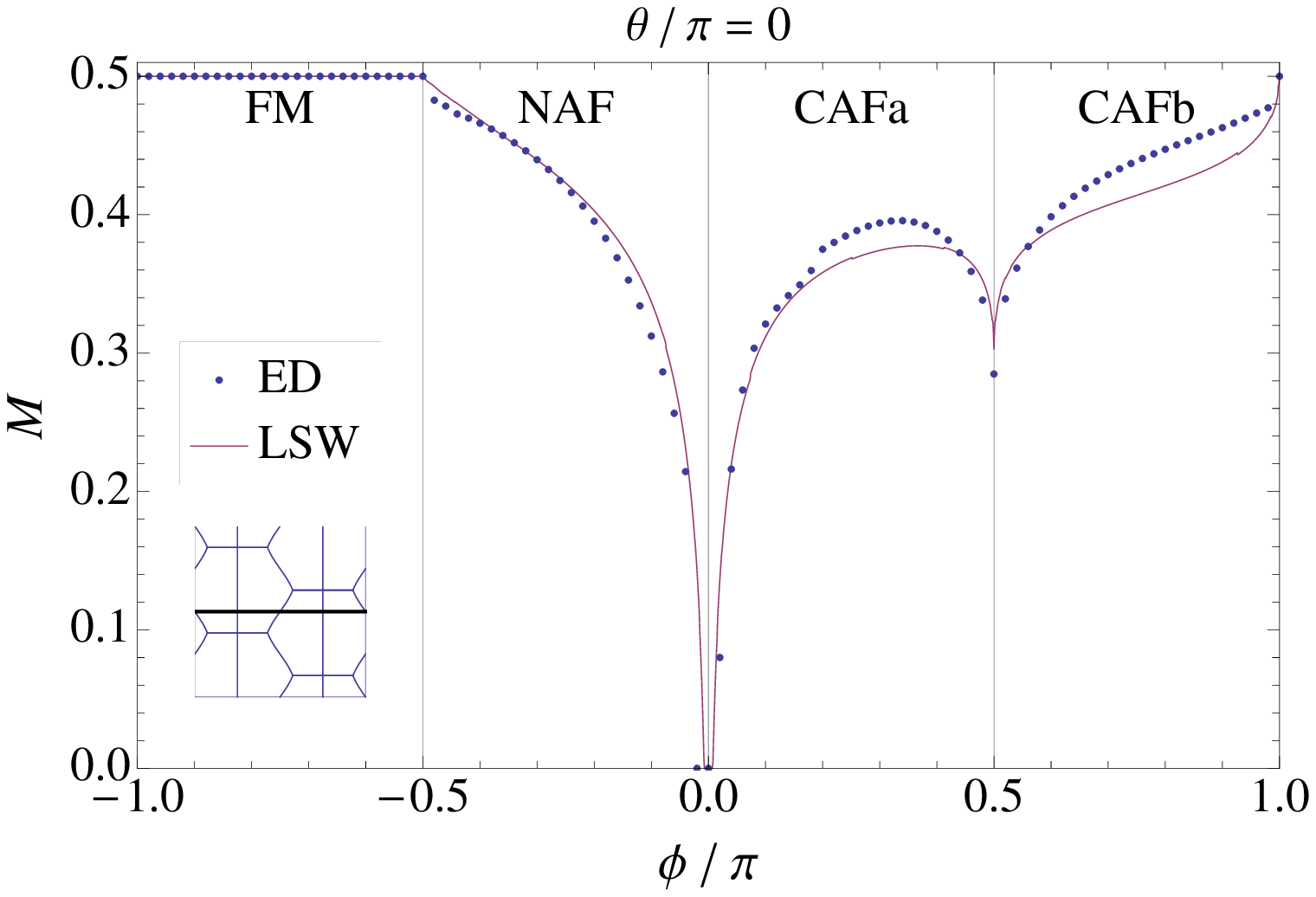}
    \caption{The extrapolated ordered moment as function of the
    frustration angle, for (top) the isotropic $\theta=\pi/4$ case and
	(bottom) the maximally anisotropic case with $\theta=0$. Those tiles
	with the maximum squareness are used for the scaling. The gray-shaded areas
	in the top plot represent the range of frustration angles $\phi$ where the
	relative error of $M^2(\vec Q)$ is above $0.1$.}
    \label{fig:moment-theta}
\end{figure}

In Fig.~\ref{fig:moment-theta} the extrapolated values for the ordered moment
$M^2(\vec Q)$ are plotted (with dots) as a function of the frustration angle
$\phi$ at two different anisotropy parameters. The top (bottom) plot corresponds
to the isotropic (maximally anisotropic) case. Again the overall agreement with
linear spin-wave theory (solid lines in Fig.~\ref{fig:moment-theta}) is very
good inside the ordered regimes of the phase diagram, and there are differences
around the borders of the columnar phases. The gray-shaded areas in the upper
plot denote the range of frustration angles $\phi$ with a relative
error $\epsilon_{\text{rel}}>0.1$ (see Eq.~(\ref{eqn:error})), indicating that
in the magnetically disordered regions the numerical results tend to become
unreliable.

In the spin-nematic region of the isotropic model, a qualitative difference
exists between linear spin wave theory and exact diagonalization: With
increasing $\phi$, the former yields a tiny region around the classical CAF/FM
phase boundary where the ordered moment vanishes before jumping to saturation in
the FM phase. In contrast, the extrapolated values for $M(\vec Q)$ from
our ED calculations remain finite at any frustration angle $\phi$.

\begin{figure}
    \includegraphics[width=\columnwidth]{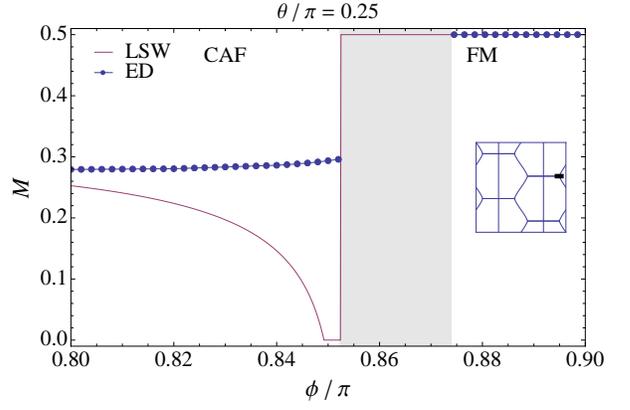}
    \caption{Ordered moment $M(\vec Q)$ as a function of the frustration
	angle, for the isotropic $\theta=\pi/4$ case around the nematic region between
	columnar and FM phases. The solid line indicates the result from linear spin
	wave theory, the dots display the scaled ED values. The gray area shows the
	region where the relative error of the extrapolated value for $M^2(\vec Q)$ is
	above 0.1.}
    \label{fig:moment-caf2fm}
\end{figure}

This behavior is displayed with greater detail in Fig.~\ref{fig:moment-caf2fm},
which shows $M(\vec Q)$ as a function of $\phi$ around the classical phase
boundary, both for linear spin-wave theory (solid line) and exact
diagonalization (dots). The extrapolated moment shows a nearly constant $\phi$
dependence $M(\vec Q_{\text{CAFa}})\approx M^{\phi=0}(\vec Q_{\text{NAF}})$ in
the columnar phase for $\phi/\pi\le0.855$, as it does in the FM phase with
$M(\vec Q_{\text{FM}})=S$ for $\phi/\pi\ge0.874$. However, for
$0.855<\phi/\pi<0.874$, $M_N(\vec Q)$ suddenly shows erratic behavior for
different tile sizes (Fig.~\ref{fig:correl-iso-caf2fm} shows this for the
long-distance correlation function), correspondingly we have
$\epsilon_{\text{rel}}>0.1$ (no significant digits for $M^2(\vec Q)$) in that
range of the frustration angle.

The point $\phi=0.874\pi$, where we can extrapolate $M_N(\vec Q)$ to a stable
limit (full polarization) again can be regarded as an upper bound of the border
between spin nematic and FM phase. According to
Fig.~\ref{fig:correl-iso-caf2fm}, the minimum value for $\phi$ where we get a
stable $M(\vec Q)=S$ decreases as a function of tile size, disregarding the
smallest eight-site tile. However the true order parameter for the spin nematic
phase is not of magnetic type, and therefore the behavior of $M(\vec Q)$ {\em
cannot\/} be used to understand the properties of this phase, in particular the
parameter range within which it exists. Instead, one would have to calculate the
spin nematic order parameter~\cite{shannon:06} in a similar way, which is beyond
the scope of the present paper.

\begin{figure}
    \includegraphics[width=\columnwidth]{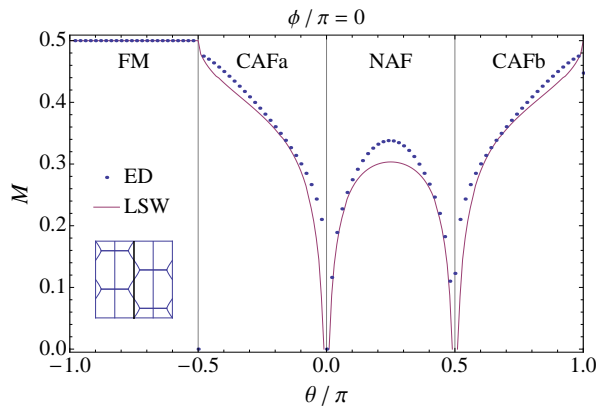}
    \caption{Ordered moment $M(\vec Q)$ for fixed frustration angle
    $\phi=0$, equivalent to $J_2=0$, i.\,e., vanishing next-nearest neighbor
    exchange. Control parameter here is the anisotropy angle $\theta$; the
    points $\theta=\pi/4$ and $\theta=-3/4\pi$ denote the isotropic case. The
    solid line denotes the result from linear spin-wave theory, the dots
    display the values extrapolated from our ED calculations.}
    \label{fig:moment-phi}
\end{figure}

Finally, the dependence of the ordered moment on the anisotropy angle $\theta$
for the unfrustrated case with only nearest neighbor interaction ($\phi=0$,
equivalent to $J_2=0$) is shown in Fig.~\ref{fig:moment-phi}. The dots display
the ED extrapolation, the solid line shows the linear spin wave result. At the
values $\theta=\pi/4$ and $\theta=-3/4\pi$, the isotropic model is recovered
with the well-known values $M(\vec Q_{\text{NAF}})\approx0.3$ and $M(\vec
Q_{\text{FM}})=S=1/2$. We point out again that the points $\theta=0$ and
$\theta=\pi/2$ at the borders of the Néel phase correspond to arrays of
independent chains. Therefore the moment suppression at these points is not a
frustration effect but a result of the effective one-dimensional character of
the model, where no magnetic moment exists at any wave vector.

\section{Summary}
\label{sect:summary}

In this work we have investigated the anisotropic frustrated square lattice
Heisenberg model using the exact diagonalization method for finite clusters
applying a finite size scaling procedure. The latter is essential to obtain
reliable values of ground state energy and ordered moment size in the
thermodynamic limit. We have also compared the numerical results with those of
linear spin wave theory~\cite{schmidt:10}. The model which we have considered is
characterized by an anisotropy parameter $\theta$ for the nearest neighbor
interaction and a frustration angle $\phi$ characterizing the ratio of next
nearest and nearest neighbor exchange.

The implementation of a stable finite size scaling procedure requires precise
criteria for the usefulness of the many possible clusters of varying size and
shape used to tile the lattice. We have introduced and described in great detail
how all possible tilings with a given area $N$ of the square or rectangular
lattice can be generated. Then we have classified the clusters according to
their spatial symmetry, compatibility with classical magnetic phases and their
geometrical compactness or squareness. We have found that a restriction to tiles
which have compatibility with classical phases and maximal squareness lead to a
very stable scaling behavior of ground state energy and ordered moment in the
region of NAF and CAF phases. Close to the classical NAF/CAF and CAF/FM
boundaries shown in Fig.~\ref{fig:phase} the relative error of energy and moment
increases and the scaling procedure becomes unstable, because a systematic
dependence of $E_{0N}$ and $M_N(\vec Q)$ on the tile area $N$ ceases to exist.
In these regions, a quantitative prediction of the size of the ordered moment
becomes difficult, if not impossible. The frustration effects of competing
exchange interactions lead to large quantum fluctuations which in turn cause
the breakdown of magnetic order.

Using the scaling results we were able to calculate the systematic dependence of
ground state energy and ordered moment as function of frustration and anisotropy
angles $\phi$ and $\theta$ respectively. We found that in the columnar phases,
introducing a spatial anisotropy strongly stabilizes the ordered moment. In fact
it becomes larger than that of the conventional unfrustrated isotropic nearest
neighbor NAF with $M(\vec Q)\approx0.3$. This stabilization effect becomes
particularly pronounced in the CAF/FM transition region where the spin nematic
phase of the isotropic model has been found.

The agreement of exact diagonalization results with spin wave calculations was
found to be generally good. Both methods predict the breakdown of magnetic order
in the transition regions at the borders of the columnar magnetic phases as
function of frustration but also in the regions where the model attains
effective quasi-one-dimensional character as function of the anisotropy for
$J_2=0$. As in earlier investigations for the isotropic model, it remains
difficult to quantify the exact size of the interval on the $\phi$ or $\theta$
axes where the ordered moment vanishes. Although not discussed here, the present
work also has given a clear framework in which the field dependence of the
ordered moment may be discussed.

\appendix

\section{All distinct eight-site tilings of the  square lattice}
\label{app:tiling}

\begin{figure*}
    \centering
    \includegraphics[width=.7\textwidth]{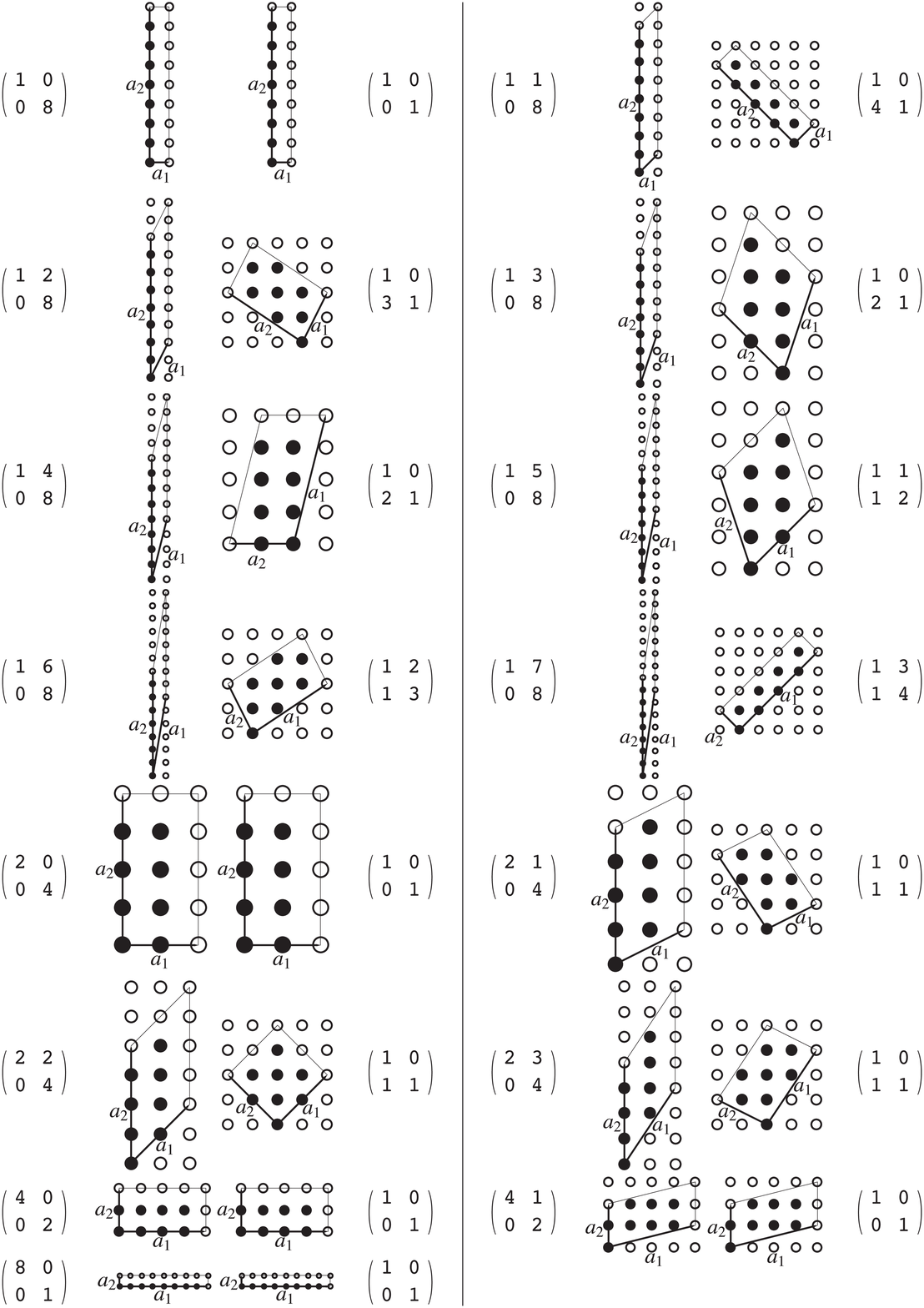}
    \caption{All 15 possible distinct square lattice tilings
    $\Lambda_{H}$ for a given tile size $N=8$.  The columns display
    the HNF matrix $H$, a graphical sketch of the HNF tile, a
    graphical sketch of the corresponding compact tile, and the
    unimodular matrix $U^{-1}$ mapping $H$ onto $M_{\text c}$.}
    \label{fig:hnf8}
\end{figure*}

To demonstrate how to find all possible tilings  with a given area $N$ of the
square lattice, Fig.~\ref{fig:hnf8} displays a list of all the distinct lattice
tilings for two-dimensional tiles of size $N=8$ created applying
Eqs.~(\ref{eqn:hnfdef}) and~(\ref{eqn:umh}).  In each row, the HNF matrix $H$ is
shown together with the HNF tile ${\cal T}_{H}$, the compact tile ${\cal
T}_{M_{\text c}}$, and the unimodular matrix $U^{-1}$ needed to map $H$ onto
$M_{\text c}$.  Note that in order to discuss orthorhombic or trigonal symmetry
breaking of the Hamiltonian, we do not regard tiles as identical which are
related by a point-group operation on the square lattice.

\bibliography{fsscaling}

\end{document}